\newif\ifMRM
\MRMfalse
\newif\ifOL
\OLfalse

\ifMRM
\documentclass[LATO1COL]{WileyNJD-v2}
\usepackage[numbers,super]{NJDnatbib}
\else
\documentclass[AMA,LATO2COL]{WileyNJD-v2}
\fi

\newcommand{\Papertitle}{Generalized Bloch model: a theory for pulsed magnetization transfer}
\articletype{Pre-Print - submitted to Magnetic Resonance in Medicine}%

\received{\today}
\revised{}
\accepted{}

\usepackage{animate}
\usepackage{tikz}
\usepackage{pgfplots}
\pgfplotsset{compat=1.17, % TODO: increase to 1.18 with newest compiler
	colormap={parula}{
			rgb255=(53,42,135)
			rgb255=(15,92,221)
			rgb255=(18,125,216)
			rgb255=(7,156,207)
			rgb255=(21,177,180)
			rgb255=(89,189,140)
			rgb255=(165,190,107)
			rgb255=(225,185,82)
			rgb255=(252,206,46)
			rgb255=(249,251,14)},
	colormap={mybluered}{
			rgb255(0cm)=(0,0,180)
			rgb255(1cm)=(0,180,180)
			rgb255(2cm)=(70,180,0)
			rgb255(3cm)=(180,180,0)
			rgb255(4cm)=(255,0,0)
			rgb255(5cm)=(128,0,0)},
	colormap={darkrainbow}{
			rgb255=(0.0,25.70808,87.75876000000001)
			rgb255=(4.0583250000000035,26.90097,83.43345)
			rgb255=(6.8056950000000045,46.824375,39.254445000000004)
			rgb255=(13.701914999999994,60.63160499999999,10.314750000000004)
			rgb255=(45.55779,80.56622999999999,2.769045000000002)
			rgb255=(98.71815000000001,110.64526499999998,8.500424999999995)
			rgb255=(146.70073499999998,134.35771499999998,18.486735000000003)
			rgb255=(163.235445,125.36972999999998,24.463425000000004)
			rgb255=(146.64310499999996,71.22022499999999,18.486989999999995)
			rgb255=(125.49417000000001,0.0,0.0)
			rgb255=(125.49417000000001,0.0,0.0)
		},
	colormap={rainbow}{
			rgb255=(120.21006,27.73533,134.38908)
			rgb255=(79.3203,29.9829,169.439595)
			rgb255=(63.93564,57.47343,196.13376)
			rgb255=(62.2404,92.11671,208.10142000000002)
			rgb255=(67.86111000000001,124.09931999999999,204.64489500000002)
			rgb255=(78.009345,149.32162499999998,188.61483)
			rgb255=(91.92240000000001,167.218545,164.65146000000001)
			rgb255=(109.60971,178.97149499999998,137.781855)
			rgb255=(130.921335,186.1296,112.37391000000001)
			rgb255=(154.951005,189.64809,91.43994)
			rgb255=(179.78469,189.360705,76.287585)
			rgb255=(202.609995,183.89529,66.511395)
			rgb255=(220.19556,171.046605,60.32382)
			rgb255=(229.75857,148.62062999999998,55.218210000000006)
			rgb255=(230.227515,115.76082,48.96357)
			rgb255=(223.917285,74.76804,40.922655000000006)
			rgb255=(218.626545,33.43203,33.69264)
		},
	colormap={turbo}{
			rgb255=(37.99,14.352,46.434)
			rgb255=(38.966,16.678,52.298)
			rgb255=(39.912,18.996,58.048)
			rgb255=(40.83,21.304000000000002,63.688)
			rgb255=(41.72,23.604,69.214)
			rgb255=(42.582,25.894000000000002,74.628)
			rgb255=(43.416,28.174,79.928)
			rgb255=(44.222,30.446,85.116)
			rgb255=(45.0,32.708,90.19200000000001)
			rgb255=(45.75,34.961999999999996,95.15599999999999)
			rgb255=(46.472,37.206,100.00800000000001)
			rgb255=(47.164,39.44,104.74600000000001)
			rgb255=(47.83,41.666,109.372)
			rgb255=(48.468,43.882,113.88400000000001)
			rgb255=(49.078,46.088,118.28399999999999)
			rgb255=(49.66,48.286,122.57199999999999)
			rgb255=(50.214000000000006,50.474,126.74799999999999)
			rgb255=(50.73800000000001,52.654,130.81199999999998)
			rgb255=(51.236000000000004,54.824,134.762)
			rgb255=(51.705999999999996,56.984,138.6)
			rgb255=(52.148,59.136,142.324)
			rgb255=(52.559999999999995,61.278,145.936)
			rgb255=(52.946000000000005,63.412,149.43599999999998)
			rgb255=(53.303999999999995,65.536,152.824)
			rgb255=(53.632000000000005,67.65,156.1)
			rgb255=(53.934000000000005,69.756,159.262)
			rgb255=(54.205999999999996,71.852,162.31199999999998)
			rgb255=(54.452,73.94,165.248)
			rgb255=(54.668000000000006,76.01599999999999,168.07399999999998)
			rgb255=(54.858,78.086,170.786)
			rgb255=(55.018,80.144,173.38400000000001)
			rgb255=(55.152,82.194,175.872)
			rgb255=(55.25600000000001,84.236,178.24599999999998)
			rgb255=(55.334,86.268,180.508)
			rgb255=(55.382,88.29,182.656)
			rgb255=(55.401999999999994,90.304,184.69400000000002)
			rgb255=(55.396,92.306,186.618)
			rgb255=(55.36,94.30199999999999,188.428)
			rgb255=(55.296,96.288,190.12800000000001)
			rgb255=(55.206,98.264,191.714)
			rgb255=(55.086,100.22999999999999,193.18800000000002)
			rgb255=(54.937999999999995,102.18799999999999,194.55)
			rgb255=(54.762,104.13799999999999,195.798)
			rgb255=(54.54599999999999,106.08,196.922)
			rgb255=(54.212,108.03,197.85999999999999)
			rgb255=(53.756,109.99000000000001,198.606)
			rgb255=(53.184,111.958,199.166)
			rgb255=(52.504,113.934,199.546)
			rgb255=(51.724000000000004,115.916,199.752)
			rgb255=(50.849999999999994,117.9,199.792)
			rgb255=(49.891999999999996,119.88600000000001,199.67)
			rgb255=(48.854,121.874,199.394)
			rgb255=(47.748000000000005,123.86200000000001,198.97)
			rgb255=(46.576,125.84599999999999,198.404)
			rgb255=(45.352,127.826,197.702)
			rgb255=(44.078,129.802,196.872)
			rgb255=(42.764,131.772,195.918)
			rgb255=(41.416,133.732,194.846)
			rgb255=(40.042,135.684,193.666)
			rgb255=(38.652,137.624,192.38)
			rgb255=(37.25,139.54999999999998,190.996)
			rgb255=(35.846000000000004,141.464,189.522)
			rgb255=(34.446,143.35999999999999,187.96200000000002)
			rgb255=(33.058,145.23999999999998,186.322)
			rgb255=(31.688,147.102,184.61)
			rgb255=(30.346,148.94400000000002,182.832)
			rgb255=(29.038000000000004,150.762,180.992)
			rgb255=(27.772000000000002,152.558,179.1)
			rgb255=(26.556,154.32999999999998,177.16)
			rgb255=(25.396,156.074,175.18)
			rgb255=(24.302,157.792,173.162)
			rgb255=(23.278,159.48,171.118)
			rgb255=(22.334,161.138,169.04999999999998)
			rgb255=(21.476,162.762,166.96800000000002)
			rgb255=(20.714,164.354,164.87400000000002)
			rgb255=(20.052,165.91,162.778)
			rgb255=(19.5,167.428,160.684)
			rgb255=(19.064,168.91,158.59799999999998)
			rgb255=(18.754,170.35,156.528)
			rgb255=(18.573999999999998,171.75,154.48)
			rgb255=(18.534,173.108,152.46)
			rgb255=(18.64,174.422,150.474)
			rgb255=(18.902,175.688,148.53)
			rgb255=(19.323999999999998,176.908,146.632)
			rgb255=(19.916,178.07999999999998,144.786)
			rgb255=(20.684,179.20000000000002,143.0)
			rgb255=(21.63,180.284,141.198)
			rgb255=(22.747999999999998,181.346,139.302)
			rgb255=(24.028,182.386,137.32)
			rgb255=(25.466,183.402,135.25400000000002)
			rgb255=(27.052,184.394,133.112)
			rgb255=(28.782000000000004,185.35999999999999,130.896)
			rgb255=(30.646,186.302,128.61599999999999)
			rgb255=(32.638,187.218,126.274)
			rgb255=(34.754000000000005,188.106,123.876)
			rgb255=(36.982,188.968,121.42599999999999)
			rgb255=(39.318,189.802,118.93199999999999)
			rgb255=(41.754000000000005,190.608,116.398)
			rgb255=(44.284,191.384,113.828)
			rgb255=(46.898,192.13,111.228)
			rgb255=(49.594,192.846,108.60600000000001)
			rgb255=(52.35999999999999,193.53,105.962)
			rgb255=(55.193999999999996,194.184,103.30600000000001)
			rgb255=(58.084,194.80599999999998,100.64200000000001)
			rgb255=(61.026,195.394,97.974)
			rgb255=(64.012,195.948,95.308)
			rgb255=(67.034,196.468,92.65)
			rgb255=(70.086,196.954,90.00399999999999)
			rgb255=(73.162,197.404,87.376)
			rgb255=(76.254,197.818,84.772)
			rgb255=(79.35600000000001,198.196,82.196)
			rgb255=(82.458,198.536,79.652)
			rgb255=(85.556,198.838,77.14999999999999)
			rgb255=(88.642,199.102,74.69)
			rgb255=(91.708,199.326,72.28)
			rgb255=(94.75,199.51000000000002,69.926)
			rgb255=(97.758,199.65599999999998,67.632)
			rgb255=(100.72399999999999,199.75799999999998,65.402)
			rgb255=(103.644,199.82,63.244)
			rgb255=(106.50999999999999,199.838,61.162000000000006)
			rgb255=(109.31599999999999,199.814,59.162000000000006)
			rgb255=(112.05199999999999,199.746,57.245999999999995)
			rgb255=(114.714,199.63400000000001,55.42399999999999)
			rgb255=(117.292,199.478,53.698)
			rgb255=(119.78200000000001,199.276,52.076)
			rgb255=(122.176,199.028,50.56)
			rgb255=(124.46600000000001,198.732,49.158)
			rgb255=(126.64599999999999,198.39,47.874)
			rgb255=(128.724,197.99800000000002,46.711999999999996)
			rgb255=(130.78799999999998,197.55,45.67)
			rgb255=(132.856,197.048,44.74)
			rgb255=(134.924,196.492,43.919999999999995)
			rgb255=(136.988,195.882,43.204)
			rgb255=(139.05,195.22,42.588)
			rgb255=(141.106,194.51,42.064)
			rgb255=(143.154,193.75,41.63)
			rgb255=(145.192,192.94,41.28)
			rgb255=(147.22,192.086,41.008)
			rgb255=(149.234,191.18599999999998,40.812)
			rgb255=(151.234,190.242,40.686)
			rgb255=(153.216,189.254,40.622)
			rgb255=(155.182,188.226,40.62)
			rgb255=(157.126,187.15800000000002,40.672000000000004)
			rgb255=(159.048,186.05,40.772000000000006)
			rgb255=(160.946,184.904,40.918)
			rgb255=(162.82000000000002,183.722,41.104)
			rgb255=(164.666,182.506,41.326)
			rgb255=(166.482,181.254,41.576)
			rgb255=(168.266,179.972,41.852000000000004)
			rgb255=(170.01999999999998,178.656,42.148)
			rgb255=(171.736,177.31,42.46)
			rgb255=(173.418,175.936,42.782)
			rgb255=(175.06,174.534,43.11)
			rgb255=(176.662,173.106,43.438)
			rgb255=(178.22400000000002,171.65200000000002,43.76)
			rgb255=(179.74,170.174,44.076)
			rgb255=(181.21,168.67399999999998,44.376)
			rgb255=(182.63400000000001,167.152,44.656)
			rgb255=(184.00799999999998,165.612,44.912)
			rgb255=(185.332,164.05,45.14)
			rgb255=(186.602,162.47199999999998,45.334)
			rgb255=(187.81799999999998,160.87800000000001,45.488)
			rgb255=(188.978,159.268,45.6)
			rgb255=(190.078,157.646,45.662000000000006)
			rgb255=(191.12,156.01,45.672000000000004)
			rgb255=(192.09799999999998,154.362,45.622)
			rgb255=(193.014,152.704,45.507999999999996)
			rgb255=(193.862,151.038,45.326)
			rgb255=(194.64600000000002,149.364,45.072)
			rgb255=(195.358,147.684,44.738)
			rgb255=(196.0,146.0,44.322)
			rgb255=(196.578,144.28,43.836000000000006)
			rgb255=(197.09799999999998,142.5,43.3)
			rgb255=(197.56199999999998,140.66,42.716)
			rgb255=(197.97199999999998,138.764,42.086)
			rgb255=(198.326,136.816,41.412)
			rgb255=(198.62800000000001,134.816,40.696)
			rgb255=(198.876,132.772,39.942)
			rgb255=(199.07,130.68200000000002,39.153999999999996)
			rgb255=(199.214,128.554,38.33)
			rgb255=(199.308,126.386,37.476)
			rgb255=(199.35,124.18599999999999,36.594)
			rgb255=(199.34400000000002,121.95400000000001,35.684)
			rgb255=(199.288,119.692,34.752)
			rgb255=(199.186,117.406,33.798)
			rgb255=(199.034,115.09799999999998,32.824)
			rgb255=(198.838,112.772,31.836)
			rgb255=(198.594,110.428,30.834)
			rgb255=(198.306,108.07199999999999,29.82)
			rgb255=(197.97400000000002,105.708,28.796)
			rgb255=(197.598,103.33399999999999,27.766000000000002)
			rgb255=(197.18,100.958,26.734)
			rgb255=(196.72,98.58200000000001,25.698)
			rgb255=(196.21599999999998,96.208,24.664)
			rgb255=(195.67399999999998,93.84,23.634)
			rgb255=(195.09,91.47999999999999,22.61)
			rgb255=(194.468,89.13,21.593999999999998)
			rgb255=(193.808,86.798,20.588)
			rgb255=(193.11,84.482,19.596)
			rgb255=(192.374,82.186,18.62)
			rgb255=(191.602,79.916,17.662)
			rgb255=(190.79600000000002,77.672,16.724)
			rgb255=(189.954,75.458,15.809999999999999)
			rgb255=(189.076,73.276,14.921999999999999)
			rgb255=(188.168,71.13199999999999,14.062)
			rgb255=(187.224,69.026,13.232)
			rgb255=(186.25,66.964,12.436)
			rgb255=(185.246,64.946,11.674)
			rgb255=(184.21,62.978,10.95)
			rgb255=(183.144,61.06,10.267999999999999)
			rgb255=(182.048,59.19799999999999,9.628)
			rgb255=(180.92600000000002,57.391999999999996,9.032)
			rgb255=(179.776,55.647999999999996,8.486)
			rgb255=(178.596,53.961999999999996,7.986)
			rgb255=(177.382,52.303999999999995,7.506)
			rgb255=(176.132,50.668,7.042)
			rgb255=(174.844,49.052,6.593999999999999)
			rgb255=(173.52,47.46,6.164)
			rgb255=(172.15800000000002,45.89,5.75)
			rgb255=(170.76,44.34,5.354)
			rgb255=(169.324,42.814,4.974)
			rgb255=(167.852,41.308,4.61)
			rgb255=(166.344,39.824,4.262)
			rgb255=(164.798,38.364,3.932)
			rgb255=(163.216,36.924,3.618)
			rgb255=(161.59799999999998,35.506,3.32)
			rgb255=(159.942,34.11,3.04)
			rgb255=(158.25,32.736,2.774)
			rgb255=(156.51999999999998,31.385999999999996,2.528)
			rgb255=(154.754,30.056,2.2960000000000003)
			rgb255=(152.952,28.748,2.0820000000000003)
			rgb255=(151.112,27.461999999999996,1.884)
			rgb255=(149.234,26.196,1.702)
			rgb255=(147.322,24.954,1.538)
			rgb255=(145.37199999999999,23.733999999999998,1.39)
			rgb255=(143.38400000000001,22.536,1.258)
			rgb255=(141.35999999999999,21.36,1.142)
			rgb255=(139.3,20.204,1.044)
			rgb255=(137.20399999999998,19.072,0.962)
			rgb255=(135.07,17.96,0.898)
			rgb255=(132.898,16.872,0.848)
			rgb255=(130.69,15.804000000000002,0.8160000000000001)
			rgb255=(128.446,14.760000000000002,0.8019999999999999)
			rgb255=(126.16400000000002,13.736,0.8019999999999999)
			rgb255=(123.84599999999999,12.734000000000002,0.8200000000000001)
			rgb255=(121.492,11.756,0.8540000000000001)
			rgb255=(119.10000000000001,10.798,0.906)
			rgb255=(116.672,9.862,0.972)
			rgb255=(114.206,8.948,1.058)
			rgb255=(111.70400000000001,8.056000000000001,1.158)
			rgb255=(109.16600000000001,7.185999999999999,1.276)
			rgb255=(106.59,6.338000000000001,1.41)
			rgb255=(103.978,5.5120000000000005,1.5599999999999998)
			rgb255=(101.328,4.707999999999999,1.7260000000000002)
			rgb255=(98.642,3.926,1.91)
			rgb255=(95.92,3.166,2.11)
		},
}
\usetikzlibrary{spy,backgrounds,arrows,decorations.pathmorphing,backgrounds,positioning,fit,matrix,calc}
\pgfdeclarelayer{background}% determine background layer
\pgfdeclarelayer{foreground}% determine foreground layer
\pgfsetlayers{background,main,foreground}% order of layers

\usepgfplotslibrary{external}
\tikzset{external/system call={/Library/TeX/texbin/pdflatex \tikzexternalcheckshellescape -halt-on-error -interaction=batchmode -jobname "\image" "\texsource"}}
\RequirePackage{shellesc}
\usepgfplotslibrary{external,fillbetween}

\usepackage{xcolor}
\definecolor{UKLred} {RGB}{207, 25,  59}
\definecolor{UKLblue}{RGB}{ 47, 63, 157}
\definecolor{NYUpurple} {RGB}{88,15,139}
\definecolor{Pastrami} {RGB}{229,85,79}
\definecolor{TheLake}{RGB}{72,159,223}
\definecolor{EastRiver}{RGB}{0,115,152}
\definecolor{SpicyMustard}{RGB}{203,160,82}
\definecolor{CentralPark}{RGB}{0,108,91}
\definecolor{ProspectPark}{RGB}{64,192,172}
\definecolor{turquois}{rgb}{0,0.75,0.75}%

\usepackage{amsmath}
\usepackage{amssymb}
\usepackage{upgreek}

\usepackage{widetext}

\usepackage{xr}

\usepackage{hyperref}
\newcommand{\linkcolor}{blue}
\hypersetup{
	pdfpagelayout=TwoPageLeft,
	pdfstartview=Fit,
	bookmarksopen=false,
	pdftitle={\Papertitle},
	pdfauthor={Jakob Assl\"ander},
	pdfsubject={Manuscript},
	breaklinks=true,
	colorlinks=true,
	linkcolor=\linkcolor,
	anchorcolor=black,
	citecolor=\linkcolor,
	filecolor=black,
	menucolor=red,
	urlcolor=\linkcolor,
	pdfencoding=auto
}

\usepackage{setspace}

\newcommand*{\citen}{}% generate error, if `\citen` is already in use
\DeclareRobustCommand*{\citen}[1]{%
	\begingroup
	\romannumeral-`\x % remove space at the beginning of \setcitestyle
	\setcitestyle{numbers}%
	\cite{#1}%
	\endgroup
}

\begin{document}

\title{\Papertitle}

\author[1,2]{Jakob Assl\"ander*}
\author[3]{Cem Gultekin}
\author[1,2]{Sebastian Flassbeck}
\author[4]{Steffen J Glaser}
\author[1,2]{Daniel K Sodickson}
\authormark{Jakob Assl\"ander}

\address[1]{\orgdiv{Center for Biomedical Imaging, Dept. of Radiology}, \orgname{New York University Grossman School of Medicine}, \orgaddress{\state{NY}, \country{USA}}}
\address[2]{\orgdiv{Center for Advanced Imaging Innovation and Research (CAI2R), Dept. of Radiology}, \orgname{New York University Grossman School of Medicine}, \orgaddress{\state{NY}, \country{USA}}}
\address[3]{\orgdiv{Courant Institute of Mathematical Sciences}, \orgname{New York University}, \orgaddress{\state{NY}, \country{USA}}}
\address[4]{\orgdiv{Department of Chemistry}, \orgname{Technische Universit\"at M\"unchen}, \orgaddress{\country{Germany}}}

\corres{*Jakob Assl\"ander, Center for Biomedical Imaging, Department of Radiology, New York University Grossman School of Medicine, 650 1st Avenue, Room 218, New York, NY 10016, USA.\\ \email{jakob.asslaender@nyumc.org}}

%\presentaddress{This is sample for present address text this is sample for present address text}
\fundingInfo{grant NIH/NIBIB R21 EB027241 and was performed under the rubric of the Center for Advanced Imaging Innovation and Research, a NIBIB Biomedical Technology Resource Center (NIH~P41 EB017183).}

\abstract[Abstract]{
	\textbf{Purpose:}
	The paper introduces a classical model to describe the dynamics of large spin-1/2 ensembles associated with nuclei bound in large molecule structures, commonly referred to as the \textit{semi-solid spin pool}, and their magnetization transfer (MT) to spins of nuclei in water. 

	\textbf{Theory and Methods:}
	Like quantum-mechanical descriptions of spin dynamics and like the original Bloch equations, but unlike existing MT models, the proposed model is based on the algebra of angular momentum in the sense that it explicitly models the rotations induced by radio-frequency (RF) pulses. It generalizes the original Bloch model to non-exponential decays, which are, e.g., observed for semi-solid spin pools. The combination of rotations with non-exponential decays is facilitated by describing the latter as Green's functions, comprised in an integro-differential equation. 

	\textbf{Results:}
	Our model describes the data of an inversion-recovery magnetization-transfer experiment with varying durations of the inversion pulse substantially better than established models. We made this observation for all measured data, but in particular for pulse durations small than 300$\upmu$s. Furthermore, we provide a linear approximation of the generalized Bloch model that reduces the simulation time by approximately a factor 15,000, enabling simulation of the spin dynamics caused by a rectangular RF-pulse in roughly 2$\upmu$s.

	\textbf{Conclusion:}
	The proposed theory unifies the original Bloch model, Henkelman's steady-state theory for magnetization transfer, and the commonly assumed rotation induced by hard pulses (i.e., strong and infinitesimally short applications of RF fields) and describes experimental data better than previous models.
}


% \jnlcitation{\cname{%
% 		\author{J. Assl\"ander}},
%         \author{Cem Gultekin},
%         \author{Sebastian Flassbeck},
%         \author{Steffen J Glaser},
%         \author{Daniel K Sodickson},
% 	(\cyear{2021}),
% 	\ctitle{\Papertitle}, \cjournal{Magnetic Resonance in Medicine}, \cvol{}.}

\maketitle

\footnotetext{\textbf{Word Count: 5169}}
\section{Introduction}
Commonly, magnetization transfer \cite{Wolff1989} is described by a 2-pool model \cite{Henkelman1993} that distinguishes between protons bound in water---the so-called \textit{free pool}---and protons bound in macromolecules, such as proteins or lipids---the so-called \textit{semi-solid pool}.
The transversal magnetization of the semi-solid pool decays rather quickly ($T_2^s \approx 10\upmu$s).
Therefore, we commonly do not observe it directly. Nevertheless, the signal or contrast in virtually all MR images is affected by these molecules \cite{Melki1992,Bieri2006} due to the indirect effect of magnetization transfer,\cite{Wolff1989} i.e. the transfer of z-magnetization between the two pools.

The original Henkelman model \cite{Henkelman1993} describes the spin-dynamics of these two pools jointly with the Bloch-McConnell equations.\cite{McConnell1958}
During a continuous radio frequency wave, a steady state evolves, which allows for solving the system of differential equations.
One of the key contributions of Henkelman et al. \cite{Henkelman1993} was to reformulate the steady-state solution of the Bloch-McConnell equations to isolate two Lorentzian lineshapes that describe the saturation of respective spin pools by the continuous RF wave.
This mathematical trick views the experiment from the perspective of absorption NMR and, in line with the theory of absorption NMR, Henkelman et al. replaced the Lorentzian with a Gaussian lineshape to more accurately describe the signal of an agar phantom,\cite{Henkelman1993} and with a super-Lorentzian lineshape to describe the signal of brain white matter.\cite{Morrison1995a}

In gases and liquids, fast molecular motion allows one to model nuclear dipole-dipole interactions as random magnetic field fluctuations \cite{Bloembergen1948} and to capture their macroscopic effect by two relaxation rates, $R_1 = 1/T_1$ and $R_2 = 1/T_2$.
This theory forms the basis for the original Bloch equations \cite{Bloch1946} and their variants, such as the Bloch-McConnell equations.\cite{McConnell1958} Equivalently, the fast molecular motion gives rise to the Lorentzian lineshape in Henkelman's RF-absorption picture.

The molecular motion of large proteins or lipids is, however, substantially slower and, as a consequence, dipole-dipole interactions cannot be approximated by random fields. Thus, the macroscopic magnetization dynamics cannot be captured by the traditional Bloch equations or a corresponding Lorentzian lineshape.
In cases where the Bloch equations fail, one can usually rely on quantum mechanics to provide an accurate model.
However, the large number of interacting nuclei in each molecule and the large number of different macromolecules present in biological tissue render a full quantum mechanical treatment infeasible for the MT effect.

Henkelman et al. were able to avoid a quantum mechanical description by focusing on a steady-state condition in the presence of continuous RF waves---a condition in which absorption NMR theory applies.
Clinical MRI, however, is performed almost exclusively with RF-pulses\cite{Hahn1950} instead of continuous RF waves, due to safety limitations associated with RF power deposition, and due to practical time constraints.
In recognition of the need for a classical pulsed-MT model, Graham et al.\cite{Graham1997} and Sled and Pike\cite{Sled2000} modified Henkelman's theory and demonstrated good agreement with experimental observations for off-resonant saturation pulses with a duration of several milliseconds. 

For short RF-pulses, however, these models fail to describe the spin dynamics accurately. This becomes apparent when approaching the extreme case of a hard pulse, i.e. a strong and infinitesimally short application of RF field. Manning et al.\cite{Manning2021} demonstrated experimentally that a hard inversion pulse does indeed invert the semi-solid pool, which stands in contrast to existing MT models that predict a saturation of the semi-solid pool to zero in this limit. 

Here, we propose a model that improves existing MT models by the following properties:
\begin{itemize}
	\setlength{\itemsep}{0pt}%
	\setlength{\parskip}{0pt}
	\setlength{\parsep}{0pt}
	\item It models the limit of a hard pulse correctly.
	\item It interpolates between the extremes of a hard pulse and a continuous wave, and the modeled spin dynamics is in line first experimental validations. 
	\item It is equivalent to the Bloch equations when assuming a Lorentzian lineshape.
\end{itemize}
It is a classical theory that generalizes the Bloch equations to arbitrary lineshapes to capture the macroscopic effect of dissipative dipole-dipole interactions in large molecule structures.

\section{Theory}
\subsection{Isolated semi-solid pool}
% A characterization of such systems often requires quantum statistics, which can be cumbersome in particular in the presence of radio-frequency (RF) magnetic fields. 
% However, it is usually possible to solve the von Neumann equation in absence of an RF-field \cite{Lowe1957}, which provides a description of the so-called free induction decay (FID) that is in most cases equivalent to the Fourier transform of the absorption lineshape \cite{Ernst1966,ERNST1974a}. And in many cases these absorption lineshapes are approximated well by analytical functions, such as Gaussian \cite{Vleck1948,Herzog1956,Henkelman1993} or super-Lorentzian lineshapes \cite{Wennerstrom1973}. 

We first derive the generalized Bloch equations for an isolated semi-solid pool before incorporating magnetization transfer to and from the free pool.

Expressing the Bloch equations\cite{Bloch1946} as
\begin{align}
	(\partial_t + R_2^s) x^s(t) & = \omega_y(t) z^s(t) \label{eq:Bloch_x}                               \\
	(\partial_t + R_2^s) y^s(t) & = -\omega_x(t) z^s(t) \label{eq:Bloch_y}                              \\
	(\partial_t + R_1^s) z^s(t) & = -\omega_y(t) x^s(t) + \omega_x(t) y^s(t) + R_1^s \label{eq:Bloch_z}
\end{align}
isolates the linear differential operators $\partial_t + R_{1,2}^s$, which are composed of the partial derivative wrt. time and relaxation.
Here, $\omega_{x,y}$ are the Rabi frequencies \cite{Rabi1938} in both dimensions.  Without loss of generality, we have described the spin system in a frame of reference that rotates at the Larmor frequency, hence $\omega_z = 0$. The three spatial components of the magnetization are denoted by $x^s(t), y^s(t), z^s(t)$ and are normalized by setting the thermal equilibrium magnetization to $z^s(0)=1$.

\begin{figure}[htbp]
	\centering
	\ifMRM
		\includegraphics[]{MT_IDE_Paper-figure0.eps}
	\else
		\ifOL
			\includegraphics[]{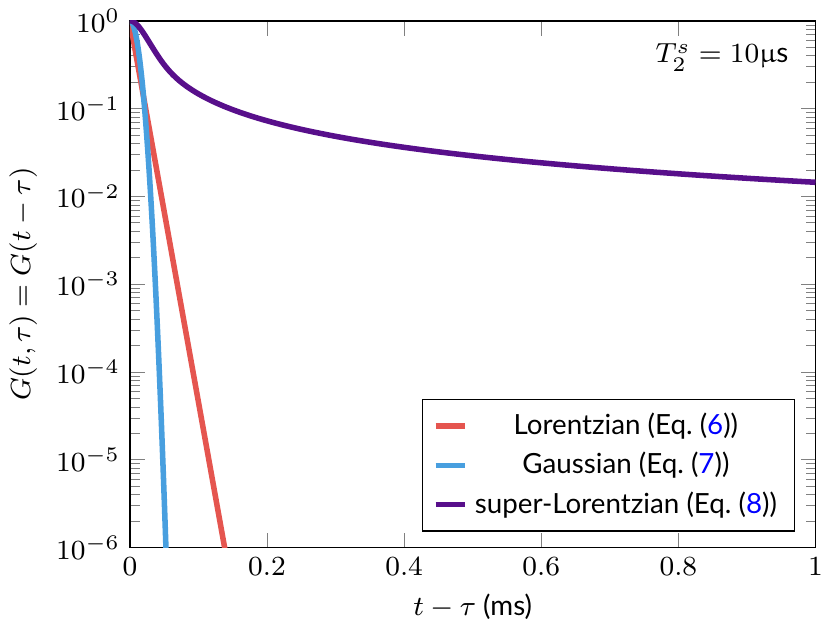}
		\else
			\begin{tikzpicture}[scale = 1]
    \begin{semilogyaxis}[
            width=\columnwidth*0.8,
            height=\textwidth*0.3,
            scale only axis,
            xmin=0,
            xmax=1,
            ymin=1e-6,
            ymax=1,
            xlabel={$t-\tau$ (ms)},
            ylabel={$G(t,\tau) = G(t-\tau)$},
            legend pos = south east,
            legend style={
                    legend image code/.code={\draw[##1,line width=1.6pt] plot coordinates {(0cm,0cm) (0.3cm,0cm)};}
                },
        ]

        \addplot [color=Pastrami,ultra thick]table[x=t_ms, y=G_Lorentzian]{Figures/Greens_Functions.txt};
        \addlegendentry{Lorentzian (Eq.~\eqref{eq:Green_Lorentzian})}

        \addplot [color=TheLake,ultra thick]table[x=t_ms, y=G_Gaussian]{Figures/Greens_Functions.txt};
        \addlegendentry{Gaussian (Eq.~\eqref{eq:Green_Gaussian})}

        \addplot [color=NYUpurple,ultra thick]table[x=t_ms, y=G_superLorentzian]{Figures/Greens_Functions.txt};
        \addlegendentry{super-Lorentzian (Eq.~\eqref{eq:Green_superLorentzian})}

		\node[anchor=north east] at (rel axis cs: 0.975, .975)  {$T_2^s = 10\upmu$s};
    \end{semilogyaxis}
\end{tikzpicture}
		\fi
	\fi
	\caption{Green's Functions. They are given by the Fourier transform of respective lineshape and are equivalent to the free induction decay of the transversal magnetization after excitation with a hard, i.e. infinitesimally short, RF-pulse (disregarding inhomogeneous broadening, i.e. assuming $T_2^{*,s}=T_2^s$).
	Despite having the same $T_2^s=10\upmu$s, the decay characteristics differ substantially: while the magnetization decays quickly when assuming a Lorentzian or Gaussian lineshape, more than 1\% of the magnetization remains after 1ms when assuming a super-Lorentzian lineshape.}
	\label{fig:Greens_Functions}
\end{figure}

Eqs.~\eqref{eq:Bloch_x} and \eqref{eq:Bloch_y} are formally solved  as follows:
\begin{align}
	x^s(t) & = \int_0^t G(t,\tau) \omega_y(\tau) z^s(\tau) d\tau \label{eq:gBloch_x}     \\
	y^s(t) & = - \int_0^t G(t,\tau) \omega_x(\tau) z^s(\tau) d\tau,  \label{eq:gBloch_y}
\end{align}
where
\begin{equation}
	G(t,\tau) = \exp (-R_2^s (t-\tau)) \;\; \forall \;\; t \geq \tau
	\label{eq:Green_Lorentzian}
\end{equation}
is the Green's function of the operator $\partial_t + R_2^s$ when assuming thermal equilibrium as initial conditions ($x^s(0) = y^s(0) = 0$, $z^s(0)=1$).
Physically, we can interpret this solution by splitting the radio-frequency field $\omega_{x,y}$ into infinitesimally short hard pulses:
Each hard pulse at the time $\tau$ excites spin coherence or transversal magnetization, which, thereafter, decays freely for the time $t-\tau$ before it is observed at the time $t$.
Since $G(t,\tau) = G(t-\tau)$ is only a function of the time between excitation and observation, Eqs. \eqref{eq:gBloch_x} and \eqref{eq:gBloch_y} describe a convolution of the Green's function (Fig.~\ref{fig:Greens_Functions})---which models an FID---with the time-varying RF-field and the evolution of the $z^s$-magnetization.

For hard pulses with an infinitesimally small flip angle, Ernst et al. showed that the FID is equivalent to the Fourier transform of the spectral absorption lineshape.\cite{Ernst1966,ERNST1974a} Therefore, we make the ansatz that we can mimic Henkelman's approach and replace the Green's function of Eq.~\eqref{eq:Green_Lorentzian} with the Fourier transform of other lineshapes, such as the Gaussian curve
\begin{equation}
	G(t,\tau) = \exp(- {R_2^s}^2 (t-\tau)^2 / 2)),
	\label{eq:Green_Gaussian}
\end{equation}
which describes the FID of gels well.\cite{Henkelman1993} Inserting this Green's function into Eqs.~\eqref{eq:gBloch_x} and \eqref{eq:gBloch_y} facilitates the description of the spin dynamics in gels during a finite RF-pulse.

As shown by Morrison et al.,\cite{Morrison1995a} the absorption lineshape of brain white matter is well-described by a super-Lorentzian lineshape.\cite{Wennerstrom1973} Hence, its Fourier transform provides us with the Green's function
% \begin{equation}
% 	G(t,\tau) = \int_0^{\pi/2} \sin \vartheta \exp \left(- {R_2^s}^2 (t - \tau)^2 \cdot  \frac{(3 \cos^2 \vartheta - 1)^2}{8} \right) d\vartheta .
% 	\label{eq:Green_superLorentzian}
% \end{equation}
\begin{equation}
	G(t,\tau) = \int_0^{1} \exp \left(- {R_2^s}^2 (t - \tau)^2 \cdot  \frac{(3 \zeta^2 - 1)^2}{8} \right) d\zeta ,
	\label{eq:Green_superLorentzian}
\end{equation}
which approaches the function $G(t,\tau) = \sqrt{\frac{2\pi}{3 R_2^s (t - \tau)}}$ for $t - \tau \gg T_2^s$.
All three Green's functions are depicted in Fig.~\ref{fig:Greens_Functions}.

% The power of the generalized Bloch equations stems from the generality of Green's functions. 
% Beyond the here presented illustrative examples, the generalized Bloch equations apply to any spin-1/2 system for which we know the absorption lineshape or can derive the FID, e.g., by solving the corresponding von Neumann equation. 
% Given the FID and using it as Green's function, the generalized Bloch equations describe the spin dynamics in the presence of any arbitrary, time varying, RF-field.

The longitudinal magnetization's return to thermal equilibrium is well-described by a mono-exponential function.\cite{Stanisz2005} Hence, Eqs.~\eqref{eq:Bloch_z}-\eqref{eq:gBloch_y}, which we dub the generalized Bloch model, describe the dynamics of an isolated semi-solid pool.
As the transversal magnetization decays before we can measure it with common clinical pulse sequences, one is often interested only in its implicit effect on the longitudinal magnetization. Hence, we can insert Eqs.~\eqref{eq:gBloch_x} and \eqref{eq:gBloch_y} into Eq.~\eqref{eq:Bloch_z} to yield a single equation that describes the dynamics of the longitudinal magnetization:
\begin{equation}
	\begin{split}
		\partial_t z^s(t) = - \omega_y(t) \int_0^t G(t,\tau) \omega_y(\tau) z^s(\tau) d\tau \\
		- \omega_x(t) \int_0^t G(t,\tau) \omega_x(\tau) z^s(\tau) d\tau \\
		+ R_1^s (1 - z^s(t)) .
	\end{split}
	\label{eq:gBloch_z}
\end{equation}
This is an integro-differential equation, which can be solved numerically.

Even though Eq.~\eqref{eq:gBloch_z} describes the dynamics of $z$-magnetization without explicitly denoting transversal magnetization, we note that transverse magnetization components are implicitly baked into this equation, and the multiplication of $z^s(\tau)$ with $\omega_{x,y}(\tau)$ as well as with $\omega_{x,y}(t)$ reflects the angular momentum algebra.

\subsection{Magnetization transfer to and from the free pool}
So far, we have assumed an isolated semi-solid spin pool. In order to describe magnetization transfer, we incorporate the generalized Bloch equations into a 2-pool model. We choose this model over the more accurate 4-pool model described by Manning et al.\cite{Manning2021} because it is difficult to determine the large number of parameters of the 4-pool model in vivo due to experimental constraints. However, the generalized Bloch model can be incorporated in models with an arbitrary number of pools. 

We can describe the three spatial components of the free pool's magnetization ($x^f$, $y^f$, and $z^f$) by the original Bloch equations due to the fast molecular motion associated with the free pool.
Since the transversal magnetization of the semi-solid pool is short-lived ($R_2^s \gg R_x$, where $R_x$ denotes the exchange rate between the two pools), we can neglect coherence transfer between the two pools and it is sufficient to account for the dynamics of the z-component $z^s$.\cite{Henkelman1993}
Given a thermal equilibrium magnetization of the free and semi-solid pool of $m_0^f$ and $m_0^s$, respectively, and normalizing their sum as $m_0^f + m_0^s = 1$, we can describe the dynamics of the coupled spin system by the generalized Bloch-McConnell equations:
\begin{align}
	(\partial_t + R_2^f) x^f(t) = & \; \omega_y(t) z^f(t) \label{eq:gBlochMcC_xf}                                           \\
	(\partial_t + R_2^f) y^f(t) = & -\omega_x(t) z^f(t)                                                                     \\
	(\partial_t + R_1^f) z^f(t) = & -\omega_y(t) x^f(t) + \omega_x(t) y^f(t)                                                \\
	                              & + m_0^f R_1^f + R_x (m_0^f z^s(t) - m_0^s z^f(t)) \nonumber                             \\
	(\partial_t + R_1^s) z^s(t) = & - \omega_y(t) \int_0^t G(t,\tau) \omega_y(\tau) z^s(\tau) d\tau \label{eq:gBlochMcC_zs} \\
	                              & - \omega_x(t) \int_0^t G(t,\tau) \omega_x(\tau) z^s(\tau) d\tau \nonumber               \\
	                              & + m_0^s R_1^s  + R_x (m_0^s z^f(t) - m_0^f z^s(t)) . \nonumber
\end{align}

\section{Methods}
\subsection{Simulations}
The simulations presented here were designed to highlight the generalized Bloch model's ability to interpolate between the well-understood extrema of a hard (i.e. infinitesimally short) RF-pulse and a continuous (i.e. infinitely long) RF-wave, and to show agreements with and deviations from existing models.
For this purpose, we focus here on rectangular RF-pulses but note that the generalized Bloch model can be applied to arbitrary pulse shapes. The validity of the generalized Bloch equations for different pulse shapes is briefly touched upon in the \textit{Discussion} section.

We simulated the dynamics of an isolated semi-solid spin-pool for two scenarios:
\begin{itemize}
	\item The spin-pool's approach to a steady state assuming a continuous wave with an off-resonance frequency of $\Delta/(2\pi) = 100$Hz and an amplitude of $\omega_1/(2\pi) = 100$Hz, as well as with an amplitude of $\omega_1/(2\pi) = 1000$Hz.
	\item The spin-pool's state at the end of on-resonant RF-pulses ($\Delta = 0$) with flip angles of $\alpha = \pi/4$, $\alpha = \pi/2$, and $\alpha = \pi$ and varying pulse durations $T_\text{RF} \in [0.1\upmu\text{s}, 0.1\text{s}]$. As we are simulating rectangular pulses, their amplitude is given by $\omega_1 = \alpha/T_\text{RF}$.
\end{itemize}
Amplitude and frequency can be translated to the rotating frame of reference used in Eq.~\eqref{eq:gBloch_z} via the expressions $\omega_x = \omega_1 \cos (\Delta t)$ and $\omega_y = \omega_1 \sin (\Delta t)$.
For both of our selected scenarios we used a subset of the models described below assuming Lorentzian, Gaussian, and super-Lorentzian lineshapes, each with $T_2^s = 10\upmu$s ($R_2^s = 10^5/$s) and $R_1^s = 1/$s.

\subsubsection{Generalized Bloch model}
We solved Eq.~\eqref{eq:gBloch_z} with the \href{https://diffeq.sciml.ai/stable/}{\textit{DifferentialEquations.jl}}\cite{Rackauckas2017} package. It provides a solver for delay differential equations and we represent $z^s(\tau)$ as an interpolated history function over which we numerically integrate.

\subsubsection{Bloch model}
In order to verify the equivalence of the generalized Bloch model using a Lorentzian lineshape to the original Bloch equations (Eqs.~\eqref{eq:Bloch_x}-\eqref{eq:Bloch_z}), we performed Bloch simulations by denoting the original Bloch equations in matrix form (cf. Eq.~(1) in Ref.~\citen{Asslander2019a}) and taking the matrix exponential.

\subsubsection{Henkelman's steady-state model}
In order to demonstrate the equivalence of the generalized Bloch model with Henkelman's model when the magnetization is in a steady state, we simulated Henkelman's model for all three lineshapes mentioned earlier. As we here assume an isolated semi-solid pool, Henkelman's steady-state solution (Eq.~(8) in Ref. \citen{Henkelman1993}) simplifies to
\begin{equation}
	R_1^s \cdot (1 - z_{\text{ss}}^s) = R_\text{RF}^s z_{\text{ss}}^s,
	\label{eq:Henkelman_steady_state}
\end{equation}
where $z_{\text{ss}}^s$ denotes the steady-state magnetization.
Eq.~\eqref{eq:Henkelman_steady_state} describes the equilibrium of $T_1$ relaxation and RF-saturation with the rate
\begin{equation}
	R_\text{RF}^s = \pi \omega_1^2 g(\Delta, T_2^s),
	\label{eq:Henkelman_Rrf}
\end{equation}
where $g(\Delta, T_2^s)$ denotes the spectral absorption lineshape (Eqs.~\eqref{eq:Lineshape_Lorentzian}-\eqref{eq:Lineshape_superLorentzian} in the appendix).

\subsubsection{Graham's spectral model}
Inspired by Henkelman's steady-state model, Graham et al.\cite{Graham1997} proposed to model the dynamics of the semi-solid spin pool with an exponential function:
\begin{equation}
	\partial_t z^s(t) = -R_\text{RF}^s z^s(t) + R_1^s (1-z^s).
	\label{eq:Graham_ODE}
\end{equation}
They postulated that the average saturation rate of an RF pulse can be described by
\begin{equation}
	\langle R_{\text{RF}}^s \rangle = \pi \int_{-\infty}^{+\infty} d\Delta' S(\Delta', \Delta, \omega_1(t)) g(\Delta', T_2^s),
	\label{eq:Graham_Rrf_spectral}
\end{equation}
where $S(\Delta', \Delta, \omega_1(t))$ denotes the power spectral density, observed at the frequency $\Delta'$, resulting from an RF-pulse with the frequency $\Delta$.\cite{Stremler1982}
For the purpose of this paper, we call the model described by Eqs.~\eqref{eq:Graham_ODE} and \eqref{eq:Graham_Rrf_spectral} Graham's spectral model.  $\langle R_\text{RF}^s \rangle$ for on-resonant rectangular pulses and for all three lineshapes are derived in the appendix (Eqs.~\eqref{eq:Rrf_Graham_spectral_Lorentzian}, \eqref{eq:Rrf_Graham_spectral_Gaussian}, and \eqref{eq:Rrf_Graham_spectral_superLorentzian}).

\subsubsection{Graham's single-frequency approximation}
In Ref.~\citen{Graham1997}, Graham et al. analyzed off-resonant RF-pulses with durations of several milliseconds. They observed that a Gaussian lineshape is approximately constant throughout the spectral response of such RF-pulses (cf. Fig.~\ref{fig:Lineshapes}).
Mathematically, this is equivalent to assuming
\begin{equation}
	S(\Delta', \Delta, \omega_1) \approx \delta(\Delta' - \Delta) \int_{0}^{T_\text{RF}} d\tau \omega_1^2(\tau).
\end{equation}
Here, $\delta(x)$ denotes the Dirac-delta distribution and the subsequent integral can be derived from Parseval's theorem.
With this assumption, the average saturation rate simplifies to
\begin{equation}
	\langle R_{\text{RF}}^s \rangle = \frac{\pi g(\Delta, T_2^s)}{T_\text{RF}} \int_{0}^{T_\text{RF}} d\tau \omega_1^2(\tau).
\end{equation}
From this average saturation rate, Graham et al. heuristically deduced a time-dependent saturation rate of
\begin{equation}
	R_{\text{RF}}^s(t) = \pi \omega_1^2(t) g(\Delta, T_2^s) ,
	\label{eq:Graham_SF_Rrf}
\end{equation}
which is essentially a time-dependent version of Henkelman's steady-state saturation rate (Eq.~\eqref{eq:Henkelman_Rrf}). Note that the last step is strictly true only for a constant $\omega_1$, which is to say a rectangular RF-pulse.
For the purpose of this paper, we call the model described by Eqs.~\eqref{eq:Graham_ODE} and \eqref{eq:Graham_SF_Rrf} Graham's single-frequency approximation.

Note that Graham's single-frequency approximation is not valid for a super-Lorentzian lineshape and on-resonant RF-pulses due to the divergence of the lineshape. A more detailed spectral analysis of this approximation can be found in the appendix (Fig.~\ref{fig:Lineshapes}).

\subsubsection{Sled's model}
Sled and Pike\cite{Sled2000} used the same exponential model as Graham and proposed to describe the time-dependent saturation rate by a convolution of the squared RF-pulse amplitude with the \textit{impulse response} $G(t-\tau)$:
\begin{equation}
	\partial_t z^s(t) = \left(-\pi \int_0^t G(t-\tau) \omega_1(\tau)^2 d\tau \right) z^s(t)  + R_1^s (1-z^s).
	\label{eq:Sled_ODE}
\end{equation}
The impulse response used by Sled and Pike is equivalent to the Green's functions derived here, and is indicated using the same notation.

\subsection{Experiments}
In order to verify the generalized Bloch equations, we performed inversion recovery experiments\cite{Gochberg2003} with varying inversion-pulse durations on a 1T NMR spectrometer (Spinsolve 43MHz, Magritek, Wellington, New Zealand).

\subsubsection{NMR samples}
We built an NMR sample of thermally cross-linked bovine serum albumin (BSA), where the BSA makes up 15\% of the overall sample weight.
We mixed the BSA powder with distilled water (weight ratio 15 to 85) and stirred it at 30$^\circ$C until the BSA was fully dissolved. We filled an NMR tube with the resulting liquid and thermally cross-linked it in a water bath at approximately 90$^\circ$C for 10 minutes.

For comparison, we also built an NMR sample consisting of distilled water doped with 0.15mM MnCl\textsubscript{2}.

\subsubsection{Pulse sequence}
We implemented a selective inversion recovery (SIR) experiment similar to the one proposed by Gochberg et. al.\cite{Gochberg2003}
Starting from thermal equilibrium (ensured by a repetition time of 30s), the free pool is inverted with a rectangular $\pi$-pulse. After an inversion time $T_\text{i}$, we excite the magnetization with a narrow-band composite $\pi/2$-pulse,\cite{Wimperis1990} consisting of 12 $\pi$-pulses with the RF-phases (104.5 + 104.5)$^\circ$, (104.5 + 255.5)$^\circ$, 104.5$^\circ$, (104.5 + 255.5)$^\circ$, (104.5 + 104.5)$^\circ$, (255.5 + 104.5)$^\circ$, (255.5 + 255.5)$^\circ$, 255.5$^\circ$, (255.5 + 255.5)$^\circ$, (255.5 + 104.5)$^\circ$, 104.5$^\circ$, and 255.5$^\circ$, followed by a $\pi/2$-pulse with the zero phase.
The goal of this composite pulse is to excite only magnetization with a well-calibrated $B_1^+$-field or, equivalently, only magnetization that was, to good approximation, fully inverted by the rectangular $\pi$-pulse.

We use phase cycling to select the FID of the excitation pulse while canceling out the FID of the (imperfect) inversion pulse, as well as the spin echo of their combination. The excitation profile of the composite pulse has a small spurious imaginary part in addition to the desired large real part.\cite{Wimperis1990} We remove this imaginary part with phase cycling as well.
Together, this requires 8 repetitions of each experiment. We acquire the FID for 1.6s with a dwell time of 100$\upmu$s.

We measured 200 data points overall, with 20 different inversion times $T_\text{i} \in [3\text{ms}, 5\text{s}]$, scaled exponentially, and 10 different durations of the inversion pulse: $T_\text{RF} \in \{22.8,45.6,114,228,342,456,570,684,798,912\}\upmu$s.

\subsubsection{Data analysis}
We performed the following analyses of the inversion-recovery (IR) NMR data.
All fits were performed with the Levenberg-Marquardt algorithm and we calculated an uncertainty from the Jacobian matrix at the point of convergence. For more details, refer to the published source code.

First, we estimated $T_2^{*,f}$ by fitting a mono-exponential decay curve to the FID of the acquisition with $T_\text{RF} = 22.8\upmu$s and $T_\text{i} = 5$s.

For the remaining analyses, we considered only the first data point of each FID. 
Plotting these data points as a function of their inversion time $T_\text{i}$ results in one inversion recovery curve for each $T_\text{RF}$.
The phase of the signal is overall very stable, which allowed us to multiply each inversion recovery curve by the complex conjugate phase factor of the longest $T_\text{i}$ data point and to disregard the imaginary part, which contains mostly noise.
% We normalized the entire dataset of $T_\text{RF}$ and $T_\text{i}$ values by normalizing all data by the largest value.

We analyzed the resulting IR curves for each $T_\text{RF}$ separately by fitting to a mono-exponential function.
Further, we performed a global fit on all data for all $T_\text{i}$ and $T_\text{RF}$.
We fitted the coupled spin-pool model (Eqs.~\eqref{eq:gBlochMcC_xf}-\eqref{eq:gBlochMcC_zs}), which describes the free pool with the original Bloch model and the semi-solid pool with the generalized Bloch model, assuming a super-Lorentzian lineshape.
The super-Lorentzian lineshape for the BSA sample was confirmed with a continuous-wave saturation experiment (the analysis can be found on \url{https://jakobasslaender.github.io/MRIgeneralizedBloch.jl/v0.2.11/build_literate/Analyze_NMR_PreSat_Data}).
We repeated the fits with the same coupled spin-pool model while replacing the description of the semi-solid pool with Graham's spectral model (Eqs.~\eqref{eq:Graham_ODE},\eqref{eq:Graham_Rrf_spectral}) and Sled's model (Eq.~\eqref{eq:Sled_ODE}), respectively.

All considered coupled-spin models comprise 7 parameters: an overall scaling factor, an inversion efficiency of the free pool, defined by $-z^f(0) / z^f(+\infty)$, the macromolecular pool fraction $m_0^s$, the longitudinal relaxation rate, where we assume $R_1 = R_1^f = R_1^s$ as proposed in the literature,\cite{Gochberg2003} $T_2^f$, $T_2^s$, and the exchange rate $R_x$.
We assumed $T_2^f \approx T_2^{*,f}$ and extracted the latter from the FID for the following reason: a finite $T_2^f$-value causes a reduction of the inversion efficiency with increasing $T_\text{RF}$ and, in the presence of off-resonance, $T_2^{*,f}$ models this effect better than $T_2^f$. We note that the effect is small regardless, as $\{T_2^f, T_2^{*,f}\} \gg T_\text{RF}$.

To assess how well the models describe the observed spin dynamics, we calculate residuals: 
For each model and each $T_\text{RF}$ separately, we calculated the $\ell_2$-norm of the difference between measured and modeled signal and normalized it by the $\ell_2$-norm of the measured signal.
First, we calculate the residuals of the actual fits, i.e. by simulating the signal with each model and the biophysical parameters that were estimated with respective model. 
For Graham's and Sled's models', this has, however, the disadvantage that the poor description at short $T_\text{RF}$ negatively affects the residuals at long $T_\text{RF}$.
To overcomes this problem, we additionally calculate the residuals between the measured signal and the one simulated with each model, but with the biophysical parameters that were estimated by fitting the generalized Bloch model. 

\subsection{Linear approximation of the generalized Bloch equations}
Solving Eqs.~\eqref{eq:gBlochMcC_xf}-\eqref{eq:gBlochMcC_zs} is numerically challenging.
For scenarios that require the simulation of many RF-pulses, as well as a repeated simulation of these RF-pulses (e.g., in a fitting routine for many voxels), the simulation time becomes impractically long.
This problem can be overcome with a linear approximation of the generalized Bloch model.
We hypothesize that we can approximate the spin dynamics using a simplified Bloch-McConnell approach \cite{McConnell1958}

\ifMRM
\else
	\pagebreak
\fi
\begin{widetext}
	\begin{equation}
		\partial_t \begin{pmatrix} x^f \\ y^f \\ z^f \\ x^s \\ z^s \\ 1 \end{pmatrix} =
		\begin{pmatrix}
			-R_2^f    & -\omega_z & \omega_y                    & 0                                    & 0                           & 0           \\
			\omega_z  & -R_2^f    & 0                           & 0                                    & 0                           & 0           \\
			-\omega_y & 0         & -R_1^f - R_{\text{x}} m_0^s & 0                                    & R_{\text{x}} m_0^f          & m_0^f R_1^f \\
			0         & 0         & 0                           & -R_2^{s,l}(T_\text{RF}/T_2^s,\alpha) & \omega_y                    & 0           \\
			0         & 0         & R_{\text{x}} m_0^s          & -\omega_y                            & -R_1^s - R_{\text{x}} m_0^f & m_0^s R_1^s \\
			0         & 0         & 0                           & 0                                    & 0                           & 0
		\end{pmatrix} \begin{pmatrix} x^f \\ y^f \\ z^f \\ x^s \\ z^s \\ 1 \end{pmatrix}.
		\label{eq:BlochMcC_Model}
	\end{equation}
\end{widetext}
As described above, we account only for exchange of the $z$-components as $R_2^s \gg R_\text{x}$.
Furthermore, we disregard the $y^s$-component as $R_2^s \gg \omega_z$ in the rotating frame, and we use RF-pulses aligned by convention along the $y$-axis.

Given this Bloch-McConnell model, the task at hand is to determine a linearized $R_2^{s,l}$ that minimizes the approximation error.
For this purpose, we can solve the full generalized Bloch-McConnell model (Eqs.~\eqref{eq:gBlochMcC_xf}-\eqref{eq:gBlochMcC_zs}) and numerically search for the $R_2^{s,l}$ that best approximates the magnetization vector at the end of the pulse.
In general, this magnetization vector depends on the eight parameters $\omega_y$, $\omega_z$, $m_0^s$, $R_1^f$, $R_1^s$, $R_2^f$, $R_2^s$, and $R_\text{x}$, as well as the initial magnetization, rendering a pre-computation of $R_2^{s,l}$ over this entire space infeasible.
Fortunately, the substantial difference in time scales ($R_2^s \gg \{R_\text{x}, R_1^f, R_1^s\}$) decouples the different processes.
For the approximation of $R_2^{s,l}$, this allows us to treat the semi-solid pool as isolated (Eq.~\eqref{eq:gBloch_z}) and to assume $R_1^s=0$. Under these assumptions, the magnetization at the end of an RF-pulse depends only on $T_2^s$, $\omega_y$, and $T_\text{RF}$.
In fact, it only depends on the ratio $T_\text{RF}/T_2^s$ and the product $\omega_y T_2^s$ or, equivalently, on $T_\text{RF}/T_2^s$ and the flip angle $\alpha$. This simplification results from a transformation of Eq. \eqref{eq:gBloch_z} to a dimensionless space, i.e. it results from substituting $\tilde{t}=t/T_\text{RF}$ and $\tilde{\tau}=\tau/T_\text{RF}$ and integrating the differential equation from $0$ to $T_\text{RF}$ while assuming a constant $\omega_y$.
In summary, these approximations and reformulations reduce the problem to a two-dimensional space and we can compare the $z^s$ at the end of the RF-pulse, as modeled by Eq.~\eqref{eq:gBloch_z}, to the one modeled by
\begin{equation}
	\partial_t \begin{pmatrix} x^s \\ z^s \end{pmatrix} =
	\begin{pmatrix}
		-R_2^{s,l}(\frac{T_\text{RF}}{T_2^s},\alpha) & \omega_y \\
		-\omega_y                                    & 0        \\
	\end{pmatrix} \begin{pmatrix} x^s \\ z^s \end{pmatrix}.
	\label{eq:Bloch_McC_2D}
\end{equation}
With $T_\text{R} \gg T_2^s$, $x^s$ vanishes between RF-pulses. Hence, we can assume $x^s(0)=0$ at the beginning of each RF-pulse and we can, further, neglect $x^s$ at the end of the pulse.

These simplifications render a pre-computation of $R_2^{s,l}$ feasible.
We solved Eq.~\eqref{eq:gBloch_z} for 4096 combinations of $T_\text{RF}/T_2^s$ and $\alpha$, evenly distributed over
$T_\text{RF}/T_2^s \in [6.7, 200]$ and $\alpha \in [0, \pi]$, and numerically optimized $R_2^{s,l}$ so that the linear model (Eq.~\eqref{eq:Bloch_McC_2D}) closely approximates the $z^s$-value at the end of the pulse as calculated with the generalized Bloch model.
This numerical optimization was performed with the \href{https://github.com/JuliaNLSolvers/NLsolve.jl}{NLsolve.jl} package that implements a trust-region algorithm.\cite{Nocedal2006}
We approximated $R_2^{s,l}(T_\text{RF}/T_2^s,\alpha)$ by cubic B-spline polynomials with the \href{http://juliamath.github.io/Interpolations.jl/latest/}{Interpolations.jl} package.
Thereafter, we solved Eq.~\eqref{eq:Bloch_McC_2D} with the approximated and interpolated $R_2^{s,l}(T_\text{RF}/T_2^s,\alpha)$ by taking the matrix exponential, and compared the result during and after RF-pulses to the solution of the full generalized Bloch-McConnell model (Eqs.~\eqref{eq:gBlochMcC_xf}-\eqref{eq:gBlochMcC_zs}).

\section{Results}
\subsection{Continuous-wave simulations}
\begin{figure}[bp]
	\centering
	\ifMRM
		\includegraphics[]{MT_IDE_Paper-figure1.eps}
	\else
		\ifOL
			\includegraphics[]{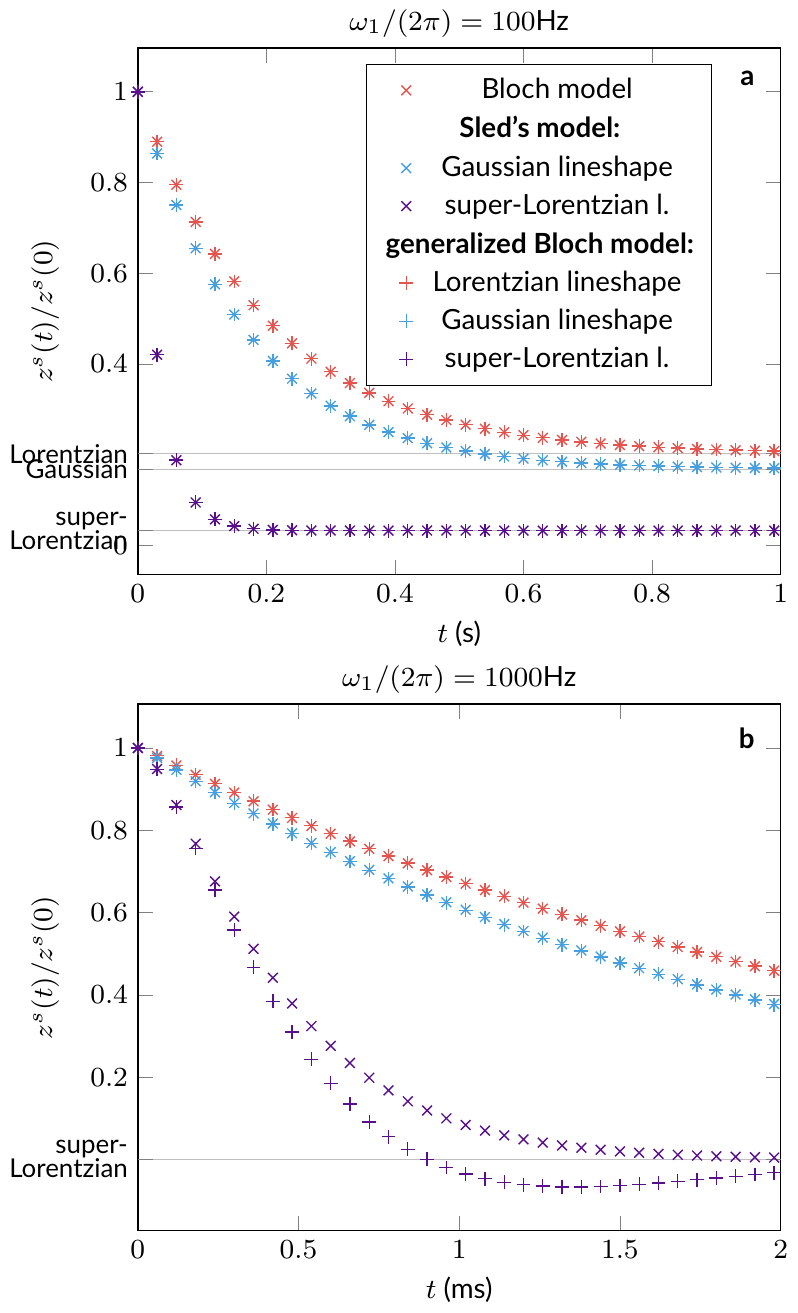}
		\else
			\begin{tikzpicture}[scale = 1]
    \begin{axis}[
            width=\columnwidth*0.75,
            height=\textwidth*0.3,
            scale only axis,
            xmin=0,
            xmax=1,
            xlabel={$t$ (s)},
            ylabel={$z^s(t)/z^s(0)$},
            ylabel style = {yshift = -0.7cm},
            legend pos = north east,
            legend style={
                    xshift = -0.5cm,
                    % legend image code/.code={\draw[##1,line width=1.6pt] plot coordinates {(0cm,0cm) (0.3cm,0cm)};}
                },
            name=version1,
            ytick = {0,.4,.6,.8,1},
            % extra y ticks={2.027e-01, 1.684e-01, 5.783e-02},
            extra y ticks={2.021e-01, 1.681e-01, 3.279e-02},
            extra tick style={grid=major},
            extra y tick labels={\footnotesize Lorentzian, \footnotesize Gaussian, \footnotesize super-\\[-1.5ex]\footnotesize Lorentzian},
            yticklabel style={align=right},
            title={$\omega_1/(2\pi) = 100$Hz},
            title style = {yshift = - 0.2cm},
        ]

        \addplot [color=Pastrami,mark=x, only marks]table[x=t_s, y=z_Bloch]{Figures//CW_SpinDynamics_v1.txt};
        \addlegendentry{Bloch model}

        % \addlegendimage{empty legend}\addlegendentry{\hspace{-0.35cm}\textbf{Graham's single-freq. approx.:}}
        \addlegendimage{empty legend}\addlegendentry{\hspace{-0.35cm}\textbf{Sled's model:}}
        % \addplot [color=Pastrami,ultra thick, densely dotted]table[x=t_s, y=z_Sled_Lorentzian]{Figures/CW_SpinDynamics_v1.txt};
        % \addlegendentry{Lorentzian}

        \addplot [color=TheLake,mark=x, only marks]table[x=t_s, y=z_Sled_Gaussian]{Figures/CW_SpinDynamics_v1.txt};
        \addlegendentry{Gaussian lineshape}

        \addplot [color=NYUpurple,mark=x, only marks]table[x=t_s, y=z_Sled_superLorentzian]{Figures/CW_SpinDynamics_v1.txt};
        \addlegendentry{super-Lorentzian l.}

        \addlegendimage{empty legend}\addlegendentry{\hspace{-0.35cm}\textbf{generalized Bloch model:}}
        \addplot [color=Pastrami,mark=+, only marks]table[x=t_s, y=z_gBloch_Lorentzian]{Figures/CW_SpinDynamics_v1.txt};
        \addlegendentry{Lorentzian lineshape}

        \addplot [color=TheLake,mark=+, only marks]table[x=t_s, y=z_gBloch_Gaussian]{Figures/CW_SpinDynamics_v1.txt};
        \addlegendentry{Gaussian lineshape}

        \addplot [color=NYUpurple,mark=+, only marks]table[x=t_s, y=z_gBloch_superLorentzian]{Figures/CW_SpinDynamics_v1.txt};
        \addlegendentry{super-Lorentzian l.}

		\node[anchor=north east] at (rel axis cs: 0.975, .975)  {\textbf{a}};
    \end{axis}

    \begin{axis}[
            width=\columnwidth*0.75,
            height=\textwidth*0.3,
            scale only axis,
            xmin=0,
            xmax=2,
            xlabel={$t$ (ms)},
            ylabel={$z^s(t)/z^s(0)$},
            ylabel style = {yshift = -0.7cm},
            ytick = {0.2,.4,.6,.8,1},
            extra y ticks={0},
            extra tick style={grid=major},
            % extra y tick labels={\footnotesize steady-state},
            extra y tick labels={\footnotesize super-\\[-1.5ex]\footnotesize Lorentzian},
            yticklabel style={align=right},
            at=(version1.below south east),
            anchor=above north east,
            title={$\omega_1/(2\pi) = 1000$Hz},
            title style = {yshift = - 0.2cm},
        ]

        \addplot [color=Pastrami,mark=x, only marks]table[x=t_ms, y=z_Bloch]{Figures//CW_SpinDynamics_v2.txt};
        % \addlegendentry{Bloch}

        % \addlegendimage{empty legend}\addlegendentry{\hspace{-0.35cm}\textbf{Graham's 2nd approx.:}}
        % \addplot [color=Pastrami,very thick,densely dotted]table[x=t_ms, y=z_Sled_Lorentzian]{Figures/CW_SpinDynamics_v2.txt};
        % \addlegendentry{Lorentzian}

        \addplot [color=TheLake,mark=x, only marks]table[x=t_ms, y=z_Sled_Gaussian]{Figures/CW_SpinDynamics_v2.txt};
        % \addlegendentry{Gaussian}

        \addplot [color=NYUpurple,mark=x, only marks]table[x=t_ms, y=z_Sled_superLorentzian]{Figures/CW_SpinDynamics_v2.txt};
        % \addlegendentry{super-Lorentzian}

        % \addlegendimage{empty legend}\addlegendentry{\hspace{-0.35cm}\textbf{generalized Bloch:}}
        \addplot [color=Pastrami,mark=+, only marks]table[x=t_ms, y=z_gBloch_Lorentzian]{Figures/CW_SpinDynamics_v2.txt};
        % \addlegendentry{Lorentzian}

        \addplot [color=TheLake,mark=+, only marks]table[x=t_ms, y=z_gBloch_Gaussian]{Figures/CW_SpinDynamics_v2.txt};
        % \addlegendentry{Gaussian}

        \addplot [color=NYUpurple,mark=+, only marks]table[x=t_ms, y=z_gBloch_superLorentzian]{Figures/CW_SpinDynamics_v2.txt};
        % \addlegendentry{super-Lorentzian}

		\node[anchor=north east] at (rel axis cs: 0.975, .975)  {\textbf{b}};
    \end{axis}
\end{tikzpicture}
		\fi
	\fi
	\caption{Simulated dynamics of an isolated semi-solid pool during a continuous wave.
		As predicted by theory, we observe good agreement between the generalized Bloch (Eq.~\eqref{eq:gBloch_z}) and Henkelman's model (Eq.~\eqref{eq:Henkelman_steady_state}) for the case of a steady state, as well as between the generalized Bloch model with a Lorentzian lineshape and the original Bloch model.
		A comparison to Sled's theory (Eq.~\eqref{eq:Sled_ODE}) reveals deviations between the two models for stronger $\omega_1$ fields.
		The horizontal lines indicate the steady-state predicted by Henkelman's theory with respective lineshape.
		The simulations were performed with $T_2^s = 10\upmu$s, $R_1^s = 1/$s, and  $\Delta/(2\pi) = 100$Hz.}
	\label{fig:Continuous_Wave}
\end{figure}

A generalized-Bloch simulation of $z$-magnetization during a continuous wave reveals several of the model's properties. First, after an initial transition, the magnetization approaches a steady state that is equivalent to the one predicted by Henkelman's steady-state model (Eq.~\eqref{eq:Henkelman_steady_state}) for all three lineshapes, as visualized in Fig. \ref{fig:Continuous_Wave}a.
This alignment is expected as, for the case of a steady state, the equivalence of these two models can be shown mathematically:
given a continuous wave ($\omega_x = \omega_1 \sin (\Delta t)$ and $\omega_y = \omega_1 \cos (\Delta t)$, where $\omega_1$ is a constant Rabi frequency and $\Delta$ denotes the off-resonance frequency), and assuming a steady state ($\partial_t z = 0$), the generalized Bloch model (Eq. \eqref{eq:gBloch_z}) reduces to
\begin{equation}
	\begin{split}
		R_1^s (1 - z_{\text{ss}}^s) = & \lim_{t \rightarrow + \infty}
		z_{\text{ss}}^s \omega_1^2 \Bigg( \cos (\Delta t) \int_0^t G(t,\tau) \cos (\Delta \tau) d\tau \\
		&+ \sin (\Delta t) \int_0^t G(t,\tau) \sin (\Delta \tau) d\tau \Bigg) \\
		= & \; \pi \omega_1^2 g(\Delta, T_2^s) z_{\text{ss}}^s ,
	\end{split}
	\label{eq:gBloch_vs_Henkelman}
\end{equation}
which is equivalent to Henkelman's steady-state model (Eqs.~\eqref{eq:Henkelman_steady_state} and \eqref{eq:Henkelman_Rrf}).
Here, we used the definition of $G(t,\tau, T_2^s)$ as the Fourier transformation of $g(\Delta, T_2^s)$.

Second, Fig.~\ref{fig:Continuous_Wave} shows good agreement between the generalized Bloch model with a Lorentzian lineshape and the original Bloch model. This is also anticipated as the two models are---by virtue of our derivation---mathematically equivalent when assuming a Lorentzian lineshape. This can be shown by inserting the Green's function in Eq. \eqref{eq:Green_Lorentzian} into the generalized Bloch equations (Eqs. \eqref{eq:gBloch_x}, \eqref{eq:gBloch_y}, and \eqref{eq:gBloch_z}) and taking the derivative with respect to time. Applying Leibniz's integral rule returns the original Bloch equations (Eqs.~\eqref{eq:Bloch_x}-\eqref{eq:Bloch_z}).

Third, Fig.~\ref{fig:Continuous_Wave} compares the generalized Bloch model with Sled's model for a Gaussian and a super-Lorentzian lineshape. With an amplitude of $\omega_1/(2\pi) = 100$Hz, we observe good agreement between the models (Fig.~\ref{fig:Continuous_Wave}a). For an amplitude of $\omega_1/(2\pi) = 1000$Hz, however, we find substantial deviations between the models in the transient phase when assuming a super-Lorentzian lineshape (Fig.~\ref{fig:Continuous_Wave}b).
While Sled's model (Eq.~\eqref{eq:Sled_ODE}) describes a decay of the $z^s$-magnetization, the generalized Bloch model (Eq.~\eqref{eq:gBloch_z}) captures the angular momentum algebra by modeling the RF-induced rotation of the magnetization---similar to the Bloch model---which results here in a negative $z^s$-magnetization in the time span $t \gtrapprox 1$ms.

While deviations between Sled's models and the generalized Bloch model are present for all lineshapes, they are pronounced the most for the super-Lorentzian lineshape as the corresponding Green's function contains more long-lived components when assuming the same $T_2^s = 10\upmu$s (cf. Fig.~\ref{fig:Greens_Functions}).
A mathematical comparison of the two models (Eq.~\eqref{eq:gBloch_z} vs. Eq.~\eqref{eq:Sled_ODE}) reveals their equivalence when assuming $z^s(t) = z^s(\tau)$ and $\omega_1(\tau) = \omega_1(t)$.
When this condition is not fulfilled, however, the subtle difference between Eq.~\eqref{eq:gBloch_z} and Eq.~\eqref{eq:Sled_ODE} can evoke substantially different spin dynamics, as demonstrated by Fig.~\ref{fig:Continuous_Wave}b.
% and we can see from Fig.~\ref{fig:Continuous_Wave} that a constant $z$-magnetization is well approximated for lower $\omega_1$ (a), while this approximation is not appropriate for $\omega_1 = 1000$Hz in the particular example used in Fig.~\ref{fig:Continuous_Wave}b.

\begin{figure*}[b]
	\centering
	\ifMRM
		\includegraphics[]{MT_IDE_Paper-figure2.eps}
	\else
		\ifOL
			\includegraphics[]{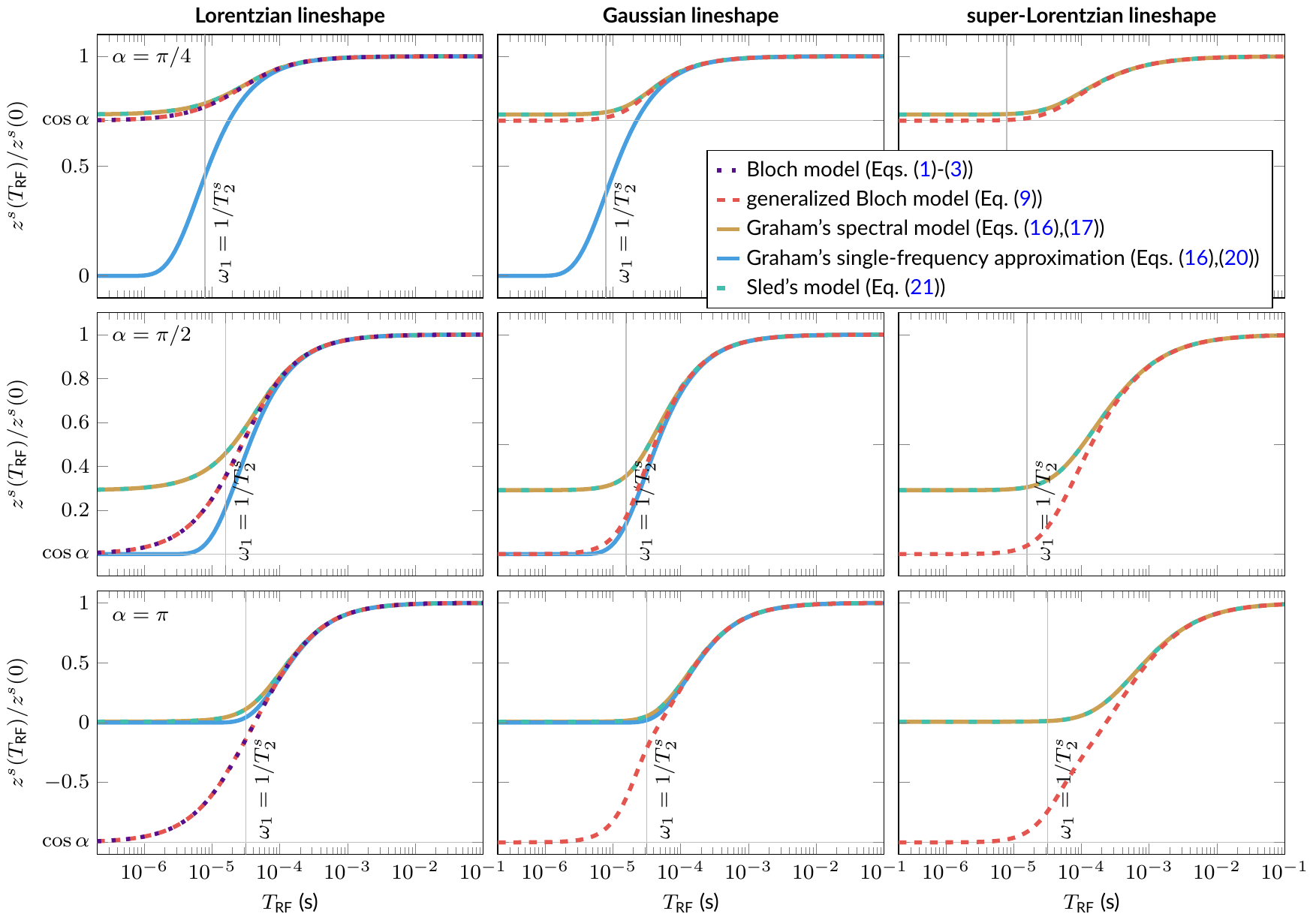}
		\else
			\input{Figures/Pulse_figure.tex}
		\fi
	\fi
	\caption{Simulated $z$-magnetization of an isolated semi-solid spin pool at the end of an RF-pulse. For a Lorentzian lineshape, we observe good agreement between the generalized Bloch and the original Bloch model (left column). For long RF-pulse durations ($T_\text{RF}$), we observe good agreement between all models for all lineshapes. However, for short RF-pulses, only the generalized Bloch model coincides with the known spin-dynamics, i.e. a rotation of the magnetization so that $z^s(T_\text{RF})/z^s(0) = \cos \alpha$, where $\alpha$ denotes the flip angle.
	For the super-Lorentzian lineshape, Graham's single frequency could not be simulated due to its singularity at $\Delta=0$.
	}
	\label{fig:Pulse}
\end{figure*}

\subsection{RF-pulse simulations}
The agreement between the generalized Bloch model with a Lorentzian lineshape and the original Bloch model is confirmed by simulations of the spin dynamics evoked by RF-pulses (left column of Fig.~\ref{fig:Pulse}). Fig.~\ref{fig:Pulse} further shows good agreement between all described models for long RF-pulse durations ($T_\text{RF} \gg T_2^s$). However, the figure also demonstrates that all established models fail to describe the expected rotation induced by short RF-pulses ($T_\text{RF} \ll T_2^s$).
In this limit the pulse approaches a hard pulse that instantaneously rotates all magnetization and we expect $z^s(T_\text{RF}) = z^s(0) \cos \alpha$. In contrast to established models, the generalized Bloch model describes such spin dynamics adequately.

The pulse duration at which the generalized Bloch model starts to deviate from Graham's and Sled's models depends on several factors.
First, deviations occur already at much longer $T_\text{RF}$ for the super-Lorentzian lineshape compared to the Lorentzian and Gaussian lineshapes, which can be explained by the slower signal decay in the corresponding Green's function (cf. Fig.~\ref{fig:Greens_Functions}) and is in line with the observations made for a continuous wave (cf. Fig.~\ref{fig:Continuous_Wave}).
Second, the deviations seem to depend on the flip angle or, more precisely, on $\omega_1$ (cf. the vertical lines in Fig.~\ref{fig:Pulse}).
This analysis indeed reveals deviations between the models for RF-pulses commonly used in clinical MRI, in particular for the super-Lorentzian lineshape: e.g., for a $\pi$-pulse with $T_\text{RF} = 1$ms, we find a deviation of $(z_\text{Sled}^s(T_\text{RF}) - z_\text{generalized Bloch}^s(T_\text{RF}))/z^s(0) \approx 0.03$. When reducing the pulse duration to $T_\text{RF} = 100\upmu$s, which brings the pulse amplitude in the range of clinical systems' maximum $B_1$-fields, we observe a deviation of $(z_\text{Sled}^s(T_\text{RF}) - z_\text{generalized Bloch}^s(T_\text{RF}))/z^s(0) \approx 0.35$.
These simulations were performed with $T_2^s = 10\upmu$s.  The results scale with the value of $T_2^s$, i.e. for longer $T_2^s$ relaxation times, the deviations become more pronounced.

\subsection{Experimental validation}
\begin{figure}[b]
	\centering
	\ifMRM
		\includegraphics[]{MT_IDE_Paper-figure3.eps}
	\else
		\ifOL
			\includegraphics[]{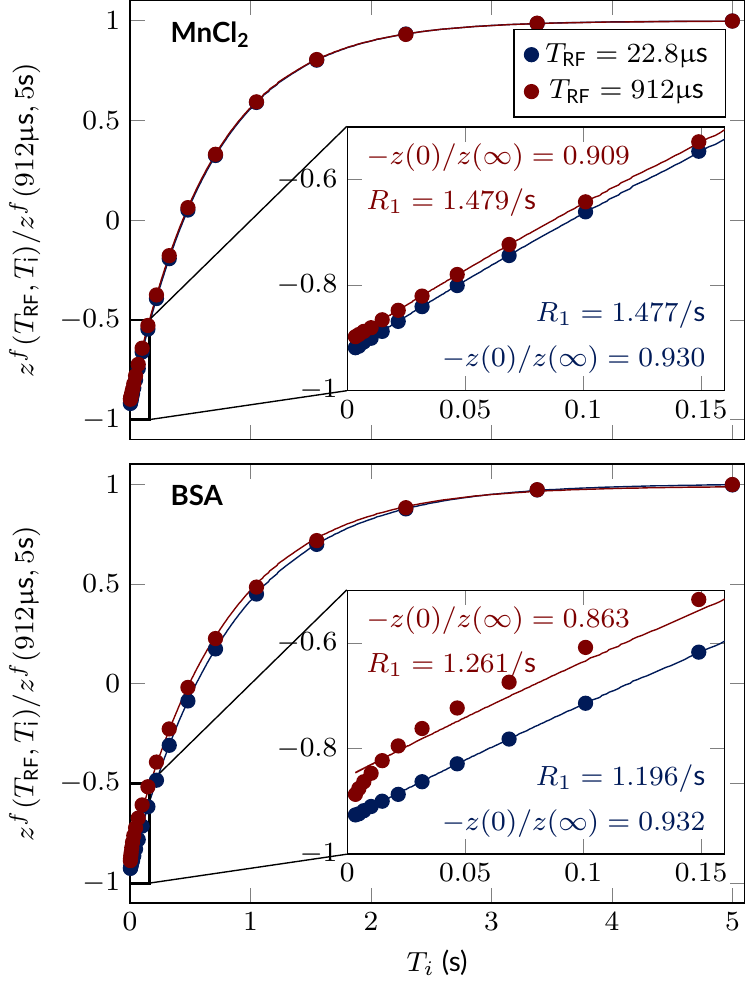}
		\else
			\input{Figures/IR_NMR_fit_MonoExp.tex}
		\fi
	\fi
	\caption{Mono-exponential fits to experimental inversion recovery data of the MnCl\textsubscript{2} and bovine serum albumin (BSA) samples with different inversion pulse durations $T_\text{RF}$ and different inversion times $T_\text{i}$. The data of the MnCl\textsubscript{2} sample is well-described by the mono-exponential model regardless of $T_\text{RF}$, as is the data of the BSA sample when inverting the magnetization with a $22.8\upmu$s pulse. Inverting the magnetization of the BSA sample with a $912\upmu$s pulse results, however, in a spin-dynamics that is clearly not mono-exponential.
	}
	\label{fig:IR_Experiment_MonoExp}
\end{figure}

\begin{figure*}[tb]
	\centering
	\ifMRM
		\includegraphics[]{MT_IDE_Paper-figure4.eps}
	\else
		\ifOL
			\includegraphics[]{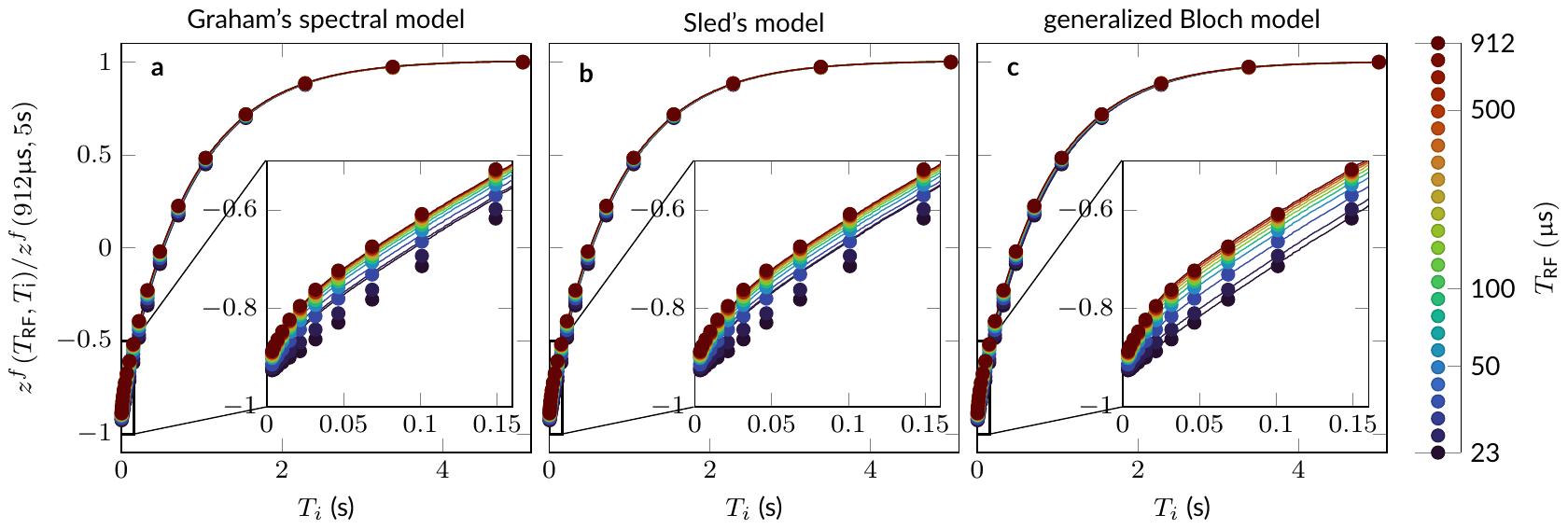}
		\else
			\input{Figures/IR_NMR_fit.tex}
		\fi
	\fi
	\caption{Inversion recovery NMR experiment of the bovine serum albumin (BSA) sample with different inversion pulse durations $T_\text{RF}$ and different inversion times $T_\text{i}$.		
	This behavior is well-described by the generalized Bloch model, and better compared to Graham's spectral model and Sled's model.
	The fitted parameters are shown in Tab.~\ref{tab:fit_parameters}.
	}
	\label{fig:IR_Experiment}
\end{figure*}

% First, we estimated $T_2^{*,f} = 74.979 \pm 0.017$ms for the MnCl\textsubscript{2} sample.
% The $\ell_2$-norm of the residual is 1.4\% of the $\ell_2$-norm of the signal vector. This small residual suggests that a mono-exponential decay fits the data well, even though the Shapiro-Wilk test failed at a significance level of 0.05, indicating that the residual is not Gaussian distributed.
% For the BSA sample, we find $T_2^{*,f} = 52.4055 \pm 0.0070$ms, with a relative residual of 0.63\%, and also a failed Shapiro-Wilk test.

First, we analyzed the IR curve of each $T_\text{RF}$ separately by fitting a mono-exponential curve (Fig. \ref{fig:IR_Experiment_MonoExp} shows the fits with the shortest and the longest $T_\text{RF}$).
The data of the MnCl\textsubscript{2} sample is well-described by the mono-exponential model, regardless of $T_\text{RF}$, which is evident by the small mean relative residual of all fits: $(0.138 \pm 0.034)\%$. 
The biggest difference between the curves is their inversion efficiency $-z(0) / z(+\infty)$, which can be explained by $T_2^f$-relaxation during the RF-pulse (cf. Fig. \ref{fig:IR_Experiment_MonoExp}). 
When fitting the BSA data with a mono-exponential model, we do see substantial differences between the individual fits:
using an inversion pulse with $T_\text{RF} = 22.8\upmu$s, the mono-exponential model fits the data well (relative residual of 0.13\%), which indicates that such short inversion pulses do indeed invert the magnetization of both spin pools (free and semi-solid) and we observe an effective $T_1$-relaxation of a fully mixed spin system, i.e. $z^f(t)/m_0^f \approx z^s(t)/m_0^s$ for the entire experiment.
In contrast, a pulse with the duration of $T_\text{RF} = 912\upmu$s inverts the magnetization of the free pool, but only saturates the one of the semi-solid pool. As a result, we observe the bi-exponential decay that is typical for such MT-experiments\cite{Gochberg2003} and the mono-exponential fit is overall poor (relative residual of 2.7\%).

% Second, we fitted the entire dataset with a global fit.
% For the MnCl\textsubscript{2} sample, we used a single-compartment Bloch model.
% The estimated relaxation rate $R_1 = (1.4787 \pm 0.0025)$/s compares well to the one estimated by the individual fits. The fit also estimated the inversion efficiency $-z(0) / z(+\infty) = 0.91882 \pm 0.00081$, confirming the effectiveness of the composite pulse to selectively excite only spins that were near-completely inverted (without the composite pulse, this inversion efficiency was substantially lower; not shown here).
% The relative residual of the fit is 0.70\%.
% All analyses and fits of the MnCl\textsubscript{2} sample can be found on \url{https://jakobasslaender.github.io/MRIgeneralizedBloch.jl/v0.2.11/build_literate/Analyze_NMR_Data/#MnCl_2-Sample}.

Second, we fitted the entire BSA dataset with a global fit using the Bloch-McConnell model, where the semi-solid pool was described with Graham's spectral, Sled's, or the generalized Bloch model.
The fits are depicted in Fig.~\ref{fig:IR_Experiment} and the fitted values are shown in Tab.~\ref{tab:fit_parameters}.
One can appreciate visually that the generalized Bloch model fits the measured data much better than Sled's and Graham's models.
This is confirmed by the relative residuals: when fitting the generalized Bloch model, the residual is smaller compared to a global Bloch fit of the MnCl\textsubscript{2} data (0.41\% vs. 0.70\%). When fitting Graham's spectral or Sled's model to the BSA data, the residual is 4-5 times larger (1.8\% and 2.0\%; cf. Tab.~\ref{tab:fit_parameters}).
Further, the parameters estimated with the generalized Bloch model (Tab.~\ref{tab:fit_parameters}) are in line with literature, while we observe unrealistically short $T_2^s$ values for Graham's and Sled's model. Yet, we note that the values are not directly comparable as different studies measured different samples with different experiments under different conditions, and analyzed the data with different biophysical models. 
% Note that the values are not directly comparable due to the different lineshape and model.
% For bovine white matter at 1.5T, $T_2^s = 10\upmu$s has been reported in Ref.~\citen{Morrison1995a}, and the same value has been reported for mouse white matter at 3T in Ref.~\citen{Stanisz2005}.
% Both studies used an off-resonant continuous-wave saturation and fitted a super-Lorentzian lineshape.
% The relaxation time of $T_2^s = 13\upmu$s estimated here with the generalized Bloch model for 15\% BSA at 1T seems realistic in this context, as does the $T_2^s = 5.0\upmu$s estimated with Sled's model. Graham's model results in an uncertainty larger than the estimated value (Tab. \ref{tab:fit_parameters}), which indicates that it does not allow for a realistic estimation of $T_2^s$ for the experimental conditions used here.

\begin{figure}[tbp]
	\centering
	\ifMRM
		\includegraphics[]{MT_IDE_Paper-figure5.eps}
	\else
		\ifOL
			\includegraphics[]{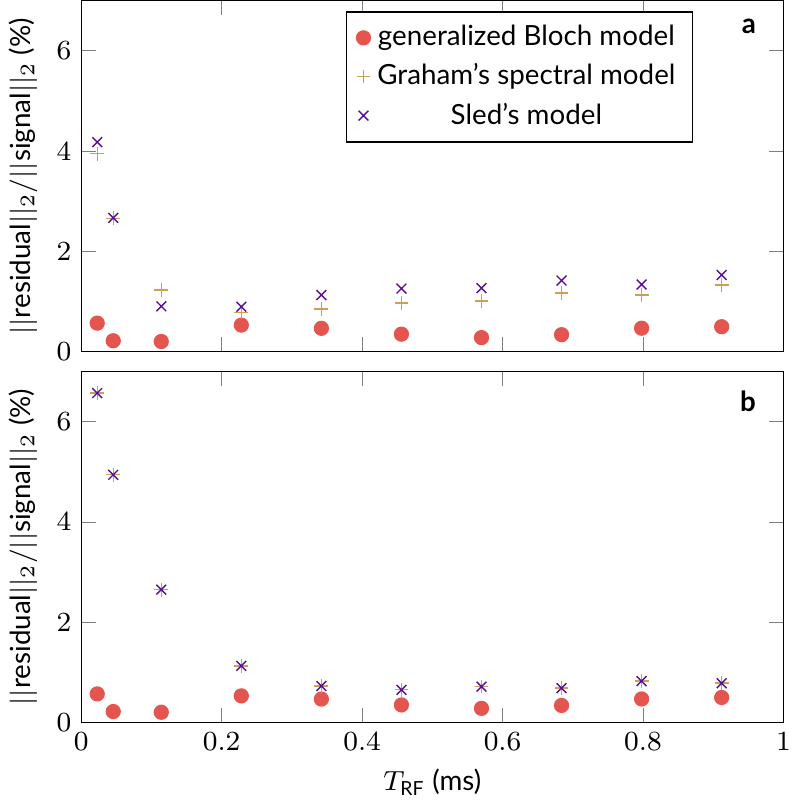}
		\else
			\begin{tikzpicture}[scale = 1]
    \begin{axis}[
            width=\textwidth*0.4,
            height=\textwidth*0.2,
            scale only axis,
            xmin=0,
            xmax=1,
            ymin=0,
            ymax=7,
            xticklabels=\empty,
            ylabel={$||\text{residual}||_2 /||\text{signal}||_2$ (\%)},
            legend entries = {generalized Bloch model, Graham's spectral model, Sled's model},
            legend style={xshift = -0.7cm},
            legend pos = north east,
            name=a,
        ]

        \addplot +[only marks, mark=*, mark options={fill=Pastrami,draw=Pastrami}] table [x=TRF_ms, y=gBloch_percent]{Figures/IR_residuum_each_fit.txt};

        \addplot +[only marks, mark=+, color=SpicyMustard] table [x=TRF_ms, y=Graham_percent]{Figures/IR_residuum_each_fit.txt};

        \addplot +[only marks, mark=x, color=NYUpurple] table [x=TRF_ms, y=Sled_percent]{Figures/IR_residuum_each_fit.txt};

		\node[anchor=north east] at (rel axis cs: .975, .975)  {\textbf{a}};
    \end{axis}

    \begin{axis}[
        width=\textwidth*0.4,
        height=\textwidth*0.2,
        scale only axis,
        xmin=0,
        xmax=1,
        ymin=0,
        ymax=7,
        xlabel={$T_\text{RF}$ (ms)},
        ylabel={$||\text{residual}||_2 /||\text{signal}||_2$ (\%)},
        at=(a.south east),
        anchor=north east,
        yshift=-0.2cm,
    ]

    \addplot +[only marks, mark=*, mark options={fill=Pastrami,draw=Pastrami}] table [x=TRF_ms, y=gBloch_percent]{Figures/IR_residuum_gBloch_est.txt};

    \addplot +[only marks, mark=+, color=SpicyMustard] table [x=TRF_ms, y=Graham_percent]{Figures/IR_residuum_gBloch_est.txt};

    \addplot +[only marks, mark=x, color=NYUpurple] table [x=TRF_ms, y=Sled_percent]{Figures/IR_residuum_gBloch_est.txt};

    \node[anchor=north east] at (rel axis cs: .975, .975)  {\textbf{b}};
\end{axis}
\end{tikzpicture}
		\fi
	\fi
	\caption{Comparison of relative residuals between different models. \textbf{a}: Residuals of the actual fits, i.e. this analysis uses the biophysical parameters of respective fit to model the signal. The disadvantage of this analysis is that residuals at long $T_\text{RF}$ are negatively affected by Graham's and Sled's models' poor description of the signal at short $T_\text{RF}$. \textbf{b}: This problem is overcome by subtracting the measured signal from signal that is simulated with the estimates from the fit with the generalized Bloch model. One can observe reduced residuals for Graham's and Sled's models at long $T_\text{RF}$. Yet, they are still substantially larger compared to the ones of the generalized Bloch model throughout, and most pronounced with an inversion pulse duration of $T_\text{RF} \lesssim 300\upmu$s.
	}
	\label{fig:IR_Residuum}
\end{figure}

Third, we used to the parameters from the generalized Bloch fit to simulate signals with all three models and we compared the norms of the residuals when subtracting such simulated signal from the measured one (Fig. \ref{fig:IR_Residuum}).
This analysis visualizes that Graham's and Sled's models describe the measured signal particularly poorly for $T_\text{RF} \lesssim 300\upmu$s. Yet, the generalized Bloch model fits the signal better for all inversion pulse durations measured here, highlighting that even small difference in the models (e.g., $(z_\text{Sled}^s(T_\text{RF}) - z_\text{generalized Bloch}^s(T_\text{RF}))/z^s(0) \approx 0.03$ for a $\pi$-pulse with $T_\text{RF} = 1$ms; cf. Fig. \ref{fig:Pulse}) can be observed experimentally. 

Additional analyses of both samples can be found on \url{https://jakobasslaender.github.io/MRIgeneralizedBloch.jl/v0.2.11/build_literate/Analyze_NMR_IR_Data}.

\begin{table*}[htbp]
	\centering
	\begin{tabular}{c|c|c|c|c}
		% \hline
		                                                        & Graham's spectral model & Sled's model        & generalized Bloch model & literature values                \\
		\hline
		$\frac{||\text{residual}||_2}{||\text{signal}||_2}(\%)$ & $1.8$                   & $2.0$               & $0.41$                  &                                  \\
		\hline
		$m_0^s$~(\%)                                            & $6.99 \pm 0.64$         & $7.15 \pm 0.38$     & $8.54 \pm 0.10$         & $7.8\pm0.4$\cite{Gochberg2003}   \\
		$T_2^s$~($\upmu$s)                                      & $2.9 \pm 5.6$           & $5.0 \pm 1.1$       & $12.95\pm0.61$          & $10.0\pm1.0$\cite{Stanisz2005}   \\
		$R_1$~(1/s)                                             & $1.179 \pm 0.067$       & $1.1782 \pm 0.0074$ & $1.1978 \pm 0.0015$     & $1.09\pm0.03$\cite{Gochberg2003} \\
		$R_x$~(1/s)                                             & $73.9 \pm 8.6$          & $60.0 \pm 8.1$      & $71.3 \pm 2.0$          & $49.2\pm5.5$\cite{Gochberg2003}  \\
	\end{tabular}
	\caption{Parameters of the bovine serum albumin (BSA) sample estimated by fitting respective model to the data. Selected parameter values reported in the literature are presented for plausibility.
		Note, however, that the values reported here for various models cannot be compared directly to the literature values, as Ref.~\citen{Gochberg2003} used an inversion recovery experiment with $T_\text{RF} = 1.5$ms, assuming a Gaussian lineshape and Graham's single-frequency model, and Ref.~\citen{Stanisz2005} used an off-resonant continuous-wave saturation experiment.
		Furthermore, we measured a 15\% solution of bovine serum albumin (BSA), thermally cross-linked, while Ref.~\citen{Gochberg2003} chemically cross-linked a 15\% (BSA) solution, and Ref.~\citen{Stanisz2005} measured mouse white matter.
		Lastly, field strengths and, likely, temperatures differed between the experiments, which could have influenced $R_1$ and $R_\text{x}$ further.
	}
	\label{tab:fit_parameters}
\end{table*}

\subsection{Linear approximation of the generalized Bloch model}
The linear approximation achieved approximately a 15,000-fold speed-up. For example, our current implementation of the full generalized Bloch model took on average 29ms for a $\pi$-pulse with $T_\text{RF} = 100\upmu$s on a single core of a 2015 desktop computer with an Intel Core i5 (I5-6500) processor.
In contrast, the linear approximation, which comprises an evaluation of the B-spline polynomials and a matrix-exponential, took on average 2.0$\upmu$s.
To give a practical example for the implications: calculating the gradient with finite differences and using an iterative fitting routine with 30 iterations for a $256^3$ matrix size, would result in a 38 days long computation on 40 cores when using the full generalized Bloch model, while the linear approximation would take only about 3.8 minutes.

\begin{figure}[tbp]
	\centering
	\ifMRM
		\includegraphics[]{MT_IDE_Paper-figure6.eps}
	\else
		\ifOL
			\includegraphics[]{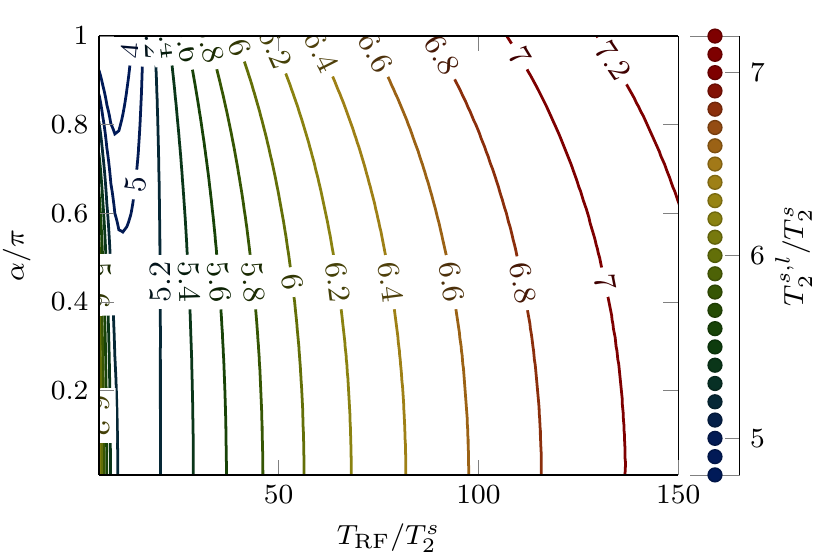}
		\else
			\begin{tikzpicture}[scale = 1]
    \begin{axis}[
        view={0}{90},
        width=\textwidth*0.33,
        height=\textwidth*0.25,
        scale only axis,
        colormap name = darkrainbow,
        xmin=5,
        xmax=150,
        xlabel={$T_\text{RF} / T_2^s$},
        ylabel={$\alpha/\pi$},
        colorbar sampled line,
        colorbar style={
            xshift = -0.5cm,
            ylabel=$T_2^{s,l} / T_2^s$,
            ylabel style = {yshift = 0.1cm},
        },
        ]

        \addplot3[contour lua={levels={4.8,5,5.2,5.4,5.6,5.8,6,6.2,6.4,6.6,6.8,7,7.2}}, thick] table [x=TRFoT2s, y=alphaopi, z=T2sloT2s]{Figures/Linearized_T2s.txt};
    \end{axis}
\end{tikzpicture}
		\fi
	\fi
	\caption{Linearized $T_2^{s,l}$ of a super-Lorentzian lineshape. The values were calculated by solving the generalized Bloch model (Eq.~\eqref{eq:gBloch_z}) and comparing the resulting $z$-magnetization to the linear Bloch-McConnell model (Eq.~\eqref{eq:Bloch_McC_2D}).
	}
	\label{fig:Linearized_T2s}
\end{figure}

Key to this speed-up is the pre-computation of the linearized $T_2^{s,l}$ as a function of $T_2^s$, $T_\text{RF}$, and $\alpha$.
For a super-Lorentzian lineshape, the pre-computed values are depicted in Fig.~\ref{fig:Linearized_T2s}. From this illustration, it becomes apparent that $T_2^{s,l}$ depends non-linearly on all three parameters. Further, we can see that the linearized $T_2^{s,l}$ is in the range of $4T_2^s-8T_2^s$, where $T_2^{s,l}$ is---by definition of the linear Bloch-McConnell equation---defined by the Lorentzian lineshape (Eq.~\eqref{eq:Lineshape_Lorentzian}) and $T_2^s$ is defined by the super-Lorentzian lineshape (Eq.~\eqref{eq:Lineshape_superLorentzian}). This observation is in line with the different decay characteristics depicted in Fig.~\ref{fig:Greens_Functions}.

\begin{figure}[tbp]
	\centering
	\ifMRM
		\includegraphics[]{MT_IDE_Paper-figure7.eps}
	\else
		\ifOL
			\includegraphics[]{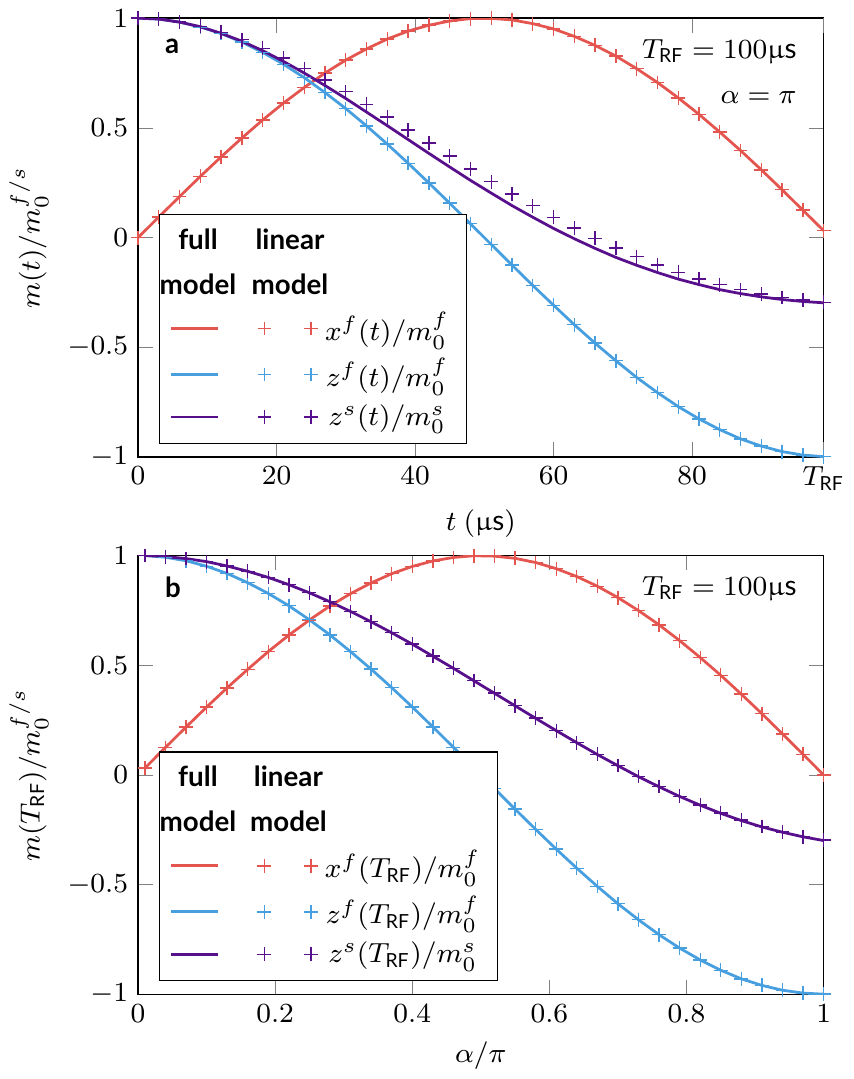}
		\else
			\begin{tikzpicture}[scale = 1]
    \begin{axis}[
        width=\columnwidth*0.8,
        height=\textwidth*0.25,
        scale only axis,
        xmin=0,
        xmax=99,
        ymin=-1,
        ymax=1,
        xlabel={$t~(\upmu\text{s})$},
        ylabel={$m(t)/m_0^{f/s}$},
        legend pos = south west,
        legend columns=2,
        legend style={
                legend image code/.code={\draw[##1] plot coordinates {(0cm,0cm) (0.475cm,0cm)};},
                cells={align=center},
            },
        extra x ticks={99},
        extra x tick labels={$T_\text{RF}$},
        name=a,
        ]

        \addlegendimage{empty legend}\addlegendentry{\hspace{-0.8cm}\textbf{full}\\ \hspace{-0.8cm}\textbf{model}}
        \addlegendimage{empty legend}\addlegendentry{\hspace{-1.375cm}\textbf{linear}\\ \hspace{-1.375cm}\textbf{model}}

        \addplot [color=Pastrami,thick]table[x=t_us, y=xf_full]{Figures/Linearized_gBloch_M_during_Pulse.txt};
        \addlegendentry{$\; \; \;$}

        \addplot [color=Pastrami,mark=+,only marks]table[x=t_us, y=xf_appx]{Figures/Linearized_gBloch_M_during_Pulse.txt};
        \addlegendentry{$x^f(t)/m_0^f$}

        \addplot [color=TheLake,thick]table[x=t_us, y=zf_full]{Figures/Linearized_gBloch_M_during_Pulse.txt};
        \addlegendentry{}

        \addplot [color=TheLake,mark=+,only marks]table[x=t_us, y=zf_appx]{Figures/Linearized_gBloch_M_during_Pulse.txt};
        \addlegendentry{$z^f(t)/m_0^f$}

        \addplot [color=NYUpurple,thick]table[x=t_us, y=zs_full]{Figures/Linearized_gBloch_M_during_Pulse.txt};
        \addlegendentry{}

        \addplot [color=NYUpurple,mark=+,only marks]table[x=t_us, y=zs_appx]{Figures/Linearized_gBloch_M_during_Pulse.txt};
        \addlegendentry{$z^s(t)/m_0^s$}

        \node[anchor=north east, align=right] at (rel axis cs: 0.975, .975)  {$T_\text{RF} = 100\upmu$s \\ $\alpha=\pi$};

		\node[anchor=north west] at (rel axis cs: 0.025, .975)  {\textbf{a}};
    \end{axis}

    \begin{axis}[
        width=\columnwidth*0.8,
        height=\textwidth*0.25,
        scale only axis,
        xmin=0,
        xmax=1,
        ymin=-1,
        ymax=1,
        xlabel={$\alpha/\pi$},
        ylabel={$m(T_\text{RF})/m_0^{f/s}$},
        legend pos = south west,
        legend columns=2,
        legend style={
            legend image code/.code={\draw[##1] plot coordinates {(0cm,0cm) (0.475cm,0cm)};},
                cells={align=center},
            },
        at=(a.south east),
        anchor=north east,
        yshift=-1.0cm,
        ]

        \addlegendimage{empty legend}\addlegendentry{\hspace{-0.8cm}\textbf{full}\\ \hspace{-0.8cm}\textbf{model}}
        \addlegendimage{empty legend}\addlegendentry{\hspace{-1.55cm}\textbf{linear}\\ \hspace{-1.55cm}\textbf{model}}

        \addplot [color=Pastrami,thick]table[x=alpha/pi, y=xf_full]{Figures/Linearized_gBloch_vary_alpha.txt};
        \addlegendentry{$\; \; \;$}

        \addplot [color=Pastrami,mark=+,only marks]table[x=alpha/pi, y=xf_appx]{Figures/Linearized_gBloch_vary_alpha.txt};
        \addlegendentry{$x^f(T_\text{RF})/m_0^f$}

        \addplot [color=TheLake,thick]table[x=alpha/pi, y=zf_full]{Figures/Linearized_gBloch_vary_alpha.txt};
        \addlegendentry{}

        \addplot [color=TheLake,mark=+,only marks]table[x=alpha/pi, y=zf_appx]{Figures/Linearized_gBloch_vary_alpha.txt};
        \addlegendentry{$z^f(T_\text{RF})/m_0^f$}

        \addplot [color=NYUpurple,thick]table[x=alpha/pi, y=zs_full]{Figures/Linearized_gBloch_vary_alpha.txt};
        \addlegendentry{}

        \addplot [color=NYUpurple,mark=+,only marks]table[x=alpha/pi, y=zs_appx]{Figures/Linearized_gBloch_vary_alpha.txt};
        \addlegendentry{$z^s(T_\text{RF})/m_0^s$}

        \node[anchor=north east] at (rel axis cs: 0.975, .975)  {$T_\text{RF} = 100\upmu$s};

		\node[anchor=north west] at (rel axis cs: 0.025, .975)  {\textbf{b}};
    \end{axis}
\end{tikzpicture}
		\fi
	\fi
	\caption{Linear approximation of the generalized Bloch model. \textbf{a}: Plotting the spin dynamics during a 100$\upmu$s $\pi$-pulse reveals deviations in $z^s$ between the generalized Bloch model and its linear approximation. The linearized relaxation time $T_2^{sl}(\alpha,T_\text{RF},T_2^s)$ was chosen such that the approximation matches the full model at the end of the RF-pulse.
	The free pool is virtually unaffected by this approximation due to the comparably slow exchange rate.
	\textbf{b}: Plotting the spin state at the end of 100$\upmu$s pulses with different flip angles reveals a good agreement between the generalized Bloch model and its linear approximation.
	}
	\label{fig:Linearized_M}
\end{figure}

A comparison of the spin dynamics during an RF-pulse, as modeled by the full generalized Bloch-McConnell equations (Eqs.~\eqref{eq:gBlochMcC_xf}-\eqref{eq:gBlochMcC_zs}), to the linear approximation (Eq.~\eqref{eq:BlochMcC_Model}) confirms that the free spin-pool is hardly affected by the approximation, but we do see deviations in the dynamics of the semi-solid pool, as expected (Fig. \ref{fig:Linearized_M}a).
However, by design, the magnetization of the semi-solid pool simulated with the two models matches at the end of the RF-pulse. This match also holds true when varying the flip angle, as shown in Fig.~\ref{fig:Linearized_M}b.
The normalized root-mean square error of the approximation, measured at the end of the RF-pulse and averaged over $\alpha \in [0,\pi]$, is $1.5 \cdot 10^{-5}$ in the $x^f$ component, $6.3 \cdot 10^{-6}$ in the $z^f$ component, and $2.7 \cdot 10^{-4}$ in the $z^s$ component.

\section{Discussion and Conclusions}
We have presented a generalization of the Bloch model to non-exponential decays or, equivalently, to non-Lorentzian lineshapes.
We derived and tested the model in the context of pulsed magnetization transfer\cite{Wolff1989,Henkelman1993,Graham1997} and demonstrated that the generalized Bloch model---in contrast to existing pulsed-MT models\cite{Graham1997,Sled2000}---unifies the original Bloch model and Henkelman's steady-state model, while also adequately describing the rotation induced by short RF-pulses, e.g., the inversion of the magnetization induced by a short $\pi$-pulse, as observed by Manning et al.\cite{Manning2021} (cf. Fig.~\ref{fig:Pulse}).

For small pulse amplitudes ($\omega_1 \ll 1/T_2^s$), the generalized Bloch model shows good agreement with Graham's and Sled's models.
With an increasing pulse amplitude, however, we find substantial deviations between the models (Fig.~\ref{fig:Pulse}) and NMR experiments confirm that the generalized Bloch model describes the spin dynamics more accurately (Fig.~\ref{fig:IR_Experiment}, \ref{fig:IR_Residuum}).

Sled's model (Eq.~\eqref{eq:Sled_ODE}) and the generalized Bloch model (Eq.~\eqref{eq:gBloch_z}) are equivalent when assuming $z^s(t) = z^s(\tau)$ and $\omega_1(\tau) = \omega_1(t)$.
Despite this resemblance, Eqs.~\eqref{eq:Sled_ODE} and \eqref{eq:gBloch_z} are mathematically quite different. As Sled's model has the form $(\partial_t + R_1) z(t) = -R_\text{RF}(t) z(t) + R_1$, it describes an exponential decay with a temporally varying decay rate.
The generalized Bloch model, on the other hand, has the form $(\partial_t + R_1) z(t) = -\omega_{y}(t) x(t) + R_1$, which describes a rotation in combination with relaxation.
This latter feature constitutes the key property of the generalized Bloch model:
explicitly modeling the rotation of the semi-solid pool's magnetization ties the magnetization transfer theory back into the algebra of angular momentum, which forms the basis of the quantum mechanical description of spin dynamics, as well as the original Bloch model.

In addition to rotations, a quantum mechanical description entails unitary evolutions that model the coherence transfer between different spins. Classical models cannot describe these dynamics to the full extent.  Instead, the generalized Bloch model is based on the assumption that such unitary transformations average to a net dissipative effect that can be captured by appropriate Green's functions.
The good agreement of the generalized Bloch model with the pulsed-MT experiment presented here, but also of Graham's and Sled's models with corresponding experiments that use long pulses with $\omega_1 \ll 1/T_2^s$, gives reason to believe that this assumption is valid for a heterogeneous mix of large molecules and most clinically feasible RF-pulses, whose amplitudes commonly vary at a time scale much larger than $T_2^s$.
A detailed analysis of the generalized Bloch model's limits of validity is beyond the scope of this paper, but one would expect that a rapid series of short pulses with large flip angles could evoke more complex behavior such as echoes that are not described by the generalized Bloch model.
Applications of the generalized Bloch model beyond magnetization transfer will also require scrutiny depending on the quantum mechanical processes one aims to approximate.

Nevertheless, we believe that the generalized Bloch model derived and validated here is the most complete theory yet formulated for pulsed-MT.
Up until now, the model arguably most commonly used for pulsed-MT is Graham's single-frequency approximation.
One of the biggest weaknesses of Graham's approximation is that the saturation rate is not well-defined for on-resonant RF-pulses when assuming a super-Lorentzian lineshape (due to its divergence at the resonant frequency).
Gochberg et al. circumvent this problem by heuristically assuming a Gaussian distribution and relying on the limited impact of the saturation on their selective inversion recovery MT approach. \cite{Gochberg2003,Gochberg2007,Dortch2011}
The bSSFP-based MT approach proposed by Gloor et al. heavily relies on the saturation of the semi-solid pool. Hence, they do use a super-Lorentzian lineshape and avoid its singularity by heuristically cutting the lineshape at 1kHz.\cite{Gloor2008}
Like Graham's spectral model and Sled's model, the generalized Bloch model does not face this issue and does not require an approximation for on-resonant RF-pulses and a super-Lorentzian lineshape.
In addition, the generalized Bloch model provides a more accurate description of the spin dynamics for short RF-pulses compared to existing models, thus, extending the well-described experimental design space.

The main disadvantage of the generalized Bloch model is its numerically challenging structure.
With this theory, we foremost aim for a ground truth to which numerically efficient approximations can be compared.
In some cases, Graham's and Sled's models might be sufficiently accurate. For others, we provided a linear approximation that allows the simulation of the dynamics during an RF-pulse in about 2$\upmu$s, rendering a broad application of the generalized Bloch model feasible.
The path to establishing a ground truth will entail further and thorough experimental validation. 

\appendix
\renewcommand{\theequation}{A.\arabic{equation}}
\renewcommand{\thefigure}{A.\arabic{figure}}

\section{Lineshapes}
Throughout this paper we use the following three lineshapes:
\begin{align}
	g^{\text{Lorentzian}}(\Delta', T_2^s) & = \frac{T_2^s}{\pi} \frac{1}{1 + (T_2^s \Delta')^2}
	\label{eq:Lineshape_Lorentzian}                                                                                                                                               \\
	g^{\text{Gaussian}}(\Delta', T_2^s)   & = \frac{T_2^s}{\sqrt{2\pi}} \exp \left( \frac{-(T_2^s \Delta')^2}{2} \right)
	\label{eq:Lineshape_Gaussian}                                                                                                                                                 \\
	% 	g^{\text{super-L.}}(\Delta')   & = T_2^s \sqrt{2/\pi} \int_0^{\pi/2} d\vartheta \sin \vartheta \frac{\exp \left(-2 \left( \frac{T_2^s \Delta'}{|3 \cos^2 \vartheta - 1|} \right)^2 \right)}{|3 \cos^2 \vartheta - 1|},
	g^{\text{super-L.}}(\Delta', T_2^s)   & = T_2^s \sqrt{2/\pi} \int_0^{1} d\zeta \frac{\exp \left(-2 \left( \frac{T_2^s \Delta'}{|3 \zeta^2 - 1|} \right)^2 \right)}{|3 \zeta^2 - 1|},
	\label{eq:Lineshape_superLorentzian}
\end{align}
which are equivalent to the ones reported in Ref. \citen{Morrison1995}.

\section{Attenuation rates in Graham's spectral model}
Here, we calculate $\langle R_\text{RF} \rangle$ for a rectangular RF-pulse with all three lineshapes defined above. We can define the spectral power density\cite{Stremler1982} for any RF-pulse $\omega_1(t)$ by
\begin{equation}
	S(\Delta', \Delta, \omega_1(t)) = \frac{1}{\sqrt{2\pi} \; T_{\text{RF}}} \left|
	\int_{-T_{\text{RF}}/2}^{T_{\text{RF}}/2} \omega_1(t) \exp(i (\Delta - \Delta') t) dt
	\right|^2 ,
	\label{eq:Power_Spectral_Density}
\end{equation}
which describes the spectral distribution of the RF-power.
For an on-resonant rectangular pulse, this Fourier integral returns a squared sinc-function:
\begin{equation}
	S(\Delta', \alpha, T_{\text{RF}}) = \frac{\alpha^2}{2\pi T_{\text{RF}}} \text{sinc}^2 \left( \frac{T_{\text{RF}} \Delta'}{2} \right) ,
	\label{eq:Power_Spectral_Density_rectangular}
\end{equation}
where we assumed an on-resonant RF-pulse ($\Delta=0$) and expressed its amplitude as a function of its flip angle $\alpha$ and its duration $T_\text{RF}$.

By inserting Eq.~\eqref{eq:Power_Spectral_Density_rectangular} and the Lorentzian lineshape (Eq.~\eqref{eq:Lineshape_Lorentzian}) in Eq.~\eqref{eq:Graham_Rrf_spectral}, we can analytically solve the integral over $\Delta'$ to calculate the saturation rate of a Lorentzian lineshape for a rectangular RF-pulse:
\begin{equation}
	% R_\text{RF}^{\text{Lorentzian}}(\alpha,T_{\text{RF}}, T_2) = \frac{\alpha^2 T_2}{T_{\text{RF}}^3} \left( T_2 \exp \left( - \frac{T_{\text{RF}}}{T_2} \right) - T_2 + T_{\text{RF}} \right) .
	R_\text{RF}^{\text{Lorentzian}}(\alpha,T_{\text{RF}}, T_2^s) = \frac{\alpha^2 T_2^s}{T_{\text{RF}}^2} \left( \frac{T_2^s}{T_\text{RF}} \exp \left( - \frac{T_{\text{RF}}}{T_2^s} \right) - \frac{T_2^s}{T_\text{RF}} + 1 \right) .
	\label{eq:Rrf_Graham_spectral_Lorentzian}
\end{equation}
For a Gaussian lineshape, this procedure results in
\begin{equation}
	\begin{split}
		R_\text{RF}^{\text{Gaussian}} &(\alpha,T_{\text{RF}}, T_2^s) = \frac{\alpha^2 T_2^s}{T_{\text{RF}}^2} \\
		&\cdot \Bigg( \frac{T_2^s}{T_{\text{RF}}} \left( \exp \left( -\frac{T_{\text{RF}}^2}{2 {T_2^s}^2} \right) - 1 \right) + \sqrt{\frac{\pi}{2}} \text{erf} \left( \frac{T_{\text{RF}}}{\sqrt{2} T_2^s} \right) \Bigg) .
	\end{split}
	\label{eq:Rrf_Graham_spectral_Gaussian}
\end{equation}
Here, $\text{erf}(x)$ denotes the error function.
For a super-Lorentzian lineshape, Fubini's theorem tells us that we can switch the order of the two integrals. Solving the integral over $\Delta'$ results in
\begin{equation}
	R_{\text{RF}}^{\text{Super-L.}}(\alpha,T_{\text{RF}}, T_2^s) = \frac{\alpha^2 T_2^s}{T_{\text{RF}}^2} f(T_{\text{RF}}/T_2^s)
	\label{eq:Rrf_Graham_spectral_superLorentzian}
\end{equation}
with
% \begin{equation}
% 	\begin{split}
% 		f(\tau) &= \int_0^{\pi/2} d\theta \frac{\sin\theta}{|3 \cos^2\theta - 1|} \\
% 		&\Bigg( \frac{4}{\tau |3 \cos^2\theta - 1|} \left( \exp \left(- \frac{\tau^2 |3 \cos^2\theta - 1|^2}{8}\right) - 1 \right) \\
% 		&+ \sqrt{2\pi} \; \text{erf} \left( \frac{\tau}{\sqrt{8}} |3 \cos^2\theta - 1| \right) \Bigg).
% 	\end{split}
% \end{equation}
\begin{equation}
	\begin{split}
		f(\tau) &= \int_0^{1} \frac{\sqrt{2\pi}}{|3 \zeta^2 - 1|} \text{erf} \left( \frac{\tau}{\sqrt{8}} |3 \zeta^2 - 1| \right) \\
		&+ \frac{4}{\tau |3 \zeta^2 - 1|^2} \left( \exp \left(- \frac{\tau^2 |3 \zeta^2 - 1|^2}{8}\right) - 1 \right) d\zeta .
	\end{split}
\end{equation}
% where we substituted $\zeta = \cos \vartheta$ and changed the integration variables accordingly.

For all three lineshapes, the saturation rate has the form $\alpha^2 T_2^s / T_{\text{RF}}^2 \cdot f(T_{\text{RF}}/T_2^s)$, where the difference between the lineshapes lies in different functions $f(\tau)$.

\section{Spectral analysis of Graham's single frequency approximation}
\begin{figure}[htbp]
	\centering
	\ifMRM
		\includegraphics[]{MT_IDE_Paper-figure8.eps}
	\else
		\ifOL
			\includegraphics[]{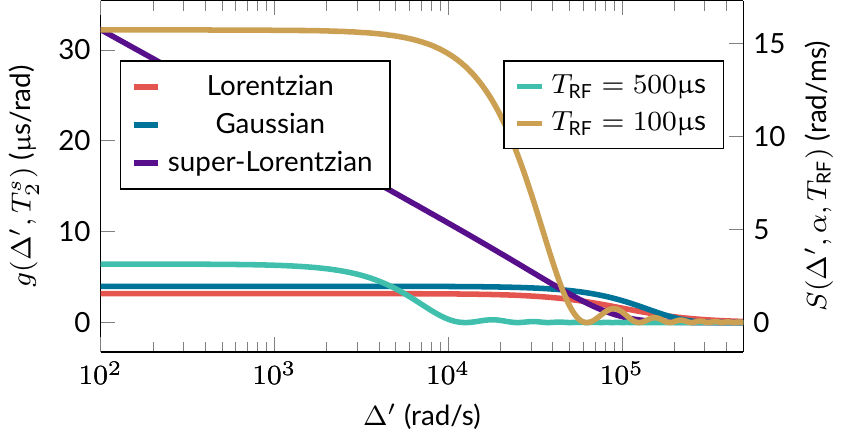}
		\else
			\begin{tikzpicture}[scale = 1]
    \begin{semilogxaxis}[
            axis y line*=left,
            width=\columnwidth*0.75,
            height=\textwidth*0.2,
            scale only axis,
            xmin=100,
            xmax=5e5,
            % ymin=0,
            % ymax=1e-5,
            xlabel={$\Delta'$ (rad/s)},
            scaled y ticks=false,
            ylabel={$g(\Delta', T_2^s)~\text{(}\upmu\text{s/rad)}$},
            ytick={0,1e-5,2e-5,3e-5},
            yticklabels={0,10,20,30},
            legend entries = {Lorentzian, Gaussian, super-Lorentzian},
            legend pos = north west,
            legend style={
                    yshift = -0.5cm,
                    legend image code/.code={\draw[##1,line width=1.6pt] plot coordinates {(0cm,0cm) (0.25cm,0cm)};}
                },
        ]

        \addplot [color=Pastrami,ultra thick]table[x=w_rad, y=D_Lorentzian]{Figures/Lineshapes.txt};
        \addplot [color=EastRiver,ultra thick]table[x=w_rad, y=D_Gaussian]{Figures/Lineshapes.txt};
        \addplot [color=NYUpurple,ultra thick]table[x=w_rad, y=D_SuperLorentzian]{Figures/Lineshapes.txt};
        % \node[minimum size=5mm, opacity=0.85, text opacity=1] at (axis cs:  21.125, 5000) {\textbf{a}};
    \end{semilogxaxis}

    \begin{semilogxaxis}[
            axis y line*=right,
            width=\columnwidth*0.75,
            height=\textwidth*0.2,
            scale only axis,
            xmin=100,
            xmax=5e5,
            % ymin=0,
            % ymax=1e-5,
            % yticklabel=\empty,
            scaled y ticks=false,
            ytick={0,5e3,1e4,1.5e4},
            yticklabels={0,5,10,15},
            ylabel={$S(\Delta',\alpha,T_\text{RF})~\text{(rad/ms)}$},
            legend entries = {$T_\text{RF} = 500\upmu$s, $T_\text{RF} = 100\upmu$s},
            legend pos = north east,
            legend style={
                    yshift = -0.5cm,
                    legend image code/.code={\draw[##1,line width=1.6pt] plot coordinates {(0cm,0cm) (0.25cm,0cm)};}
                },
        ]
        \addplot [color=ProspectPark,ultra thick]table[x=w_rad, y=PSD_500us]{Figures/Lineshapes.txt};
        \addplot [color=SpicyMustard,ultra thick]table[x=w_rad, y=PSD_100us]{Figures/Lineshapes.txt};
    \end{semilogxaxis}
\end{tikzpicture}
		\fi
	\fi
	\caption{Spectral analysis of the three lineshapes discussed in this paper and RF-pulses with different pulse durations $T_\text{RF}$.
		The Lorentzian and Gaussian lineshapes $g(\Delta', T_2^s)$ are constant throughout the spectral power density $S(\Delta',\alpha,T_\text{RF})$ of an on-resonant rectangular pulse with $T_\text{RF} = 500\upmu$s and, to a lesser degree, with $T_\text{RF} = 100\upmu$s.
		This assumption is not valid for a super-Lorentzian lineshape due to the divergence of the lineshape on-resonance.
		All lineshapes were simulated with $T_2^s = 10\upmu$s and we set $\alpha = \pi$ for both RF-pulses.}
	\label{fig:Lineshapes}
\end{figure}

In order to analyze the limitations of Graham's single frequency approximation, we plotted in Fig.~\ref{fig:Lineshapes} a Lorentzian, a Gaussian, and a super-Lorentzian lineshape along with the spectral power density functions of rectangular RF-pulses. For a Lorentzian or Gaussian lineshape in combination with an on-resonant, rectangular RF-pulse with $T_\text{RF} = 500\upmu$s, Graham's approximations appear justified: neither of those two lineshapes exhibit substantial changes throughout the spectral profile of the RF-pulse. We can, hence, approximate the lineshape by a constant value.
Reducing $T_\text{RF}$ to $100\upmu$s pushes the limits of such an approximation: the lineshapes show some variations over the larger spectral profile and we can expect some errors when using this approximation.
For a super-Lorentzian lineshape, this approximation cannot be made at all: due to the singularity on resonance ($\Delta' = 0$), the lineshape varies throughout the power spectral density, regardless of the pulse shape and duration.

\section{Acknowledgements}
We would like to thank Rafael P. Bruschweiler for fruitful discussions.

\section{Data availability statement}
In order to ensure reproducibility, we provide the following resources:
\begin{itemize}
	\item Highly efficient implementations of the generalized Bloch model and its linear approximation can be found on \url{https://github.com/JakobAsslaender/MRIgeneralizedBloch.jl}. They are written in the open source language \href{https://julialang.org}{Julia} and we registered the package "MRIgeneralizedBloch.jl," which facilitates the installation via Julia's package manager. 
	\item Scripts to reproduce all simulations, data analyses, and figures can be found in the documentation of the package:
	      \url{https://JakobAsslaender.github.io/MRIgeneralizedBloch.jl/v0.2.11}.
	      They render the code in HTML format with interactive figures and link to Jupyter notebooks that can be launched in \textit{binder}, allowing for the replication and modification of all simulations in a browser without any local installations. 
	\item All NMR data are stored in a separate github repository \url{https://github.com/JakobAsslaender/MRIgeneralizedBloch_NMRData} and are loaded and analyzed by above mentioned scripts. 
\end{itemize}

All figures and tables in this paper were created with the version 0.2.11 (commit hash b185969) of the MRIgeneralizedBloch.jl package and match the version's documentation. The commit hash of the data used for the preparation of this paper is 77950c4. 

\bibliography{library}
\end{document}